\documentclass[11pt,a4paper,epsf,epsfig,psfrag]{article}
\usepackage{jheppub}
\usepackage{amsmath,epsfig}
\usepackage{subfigure}
\usepackage{amssymb,amsfonts}
\usepackage{latexsym}
\usepackage{epsfig}
\newbox\pippobox
\def\be{\begin{equation}}
\def\ee{\end{equation}}
\def\bea{\begin{eqnarray}}
\def\eea{\end{eqnarray}}

\def\ee           {{\rm e}}

\newcommand{\beq}{\begin{equation}}
\newcommand{\eeq}{\end{equation}}
\newcommand{\beqa}{\begin{eqnarray}}
\newcommand{\eeqa}{\end{eqnarray}}
\newcommand{\beqar}{\begin{eqnarray*}}
\newcommand{\eeqar}{\end{eqnarray*}}

\renewcommand{\eqref}[1]{(\ref{#1})}

\catcode`\@=12

\title{Entanglement Temperature in Non-conformal Cases}
\author[a]{Song He,}
\author[b]{Danning Li,}
\author[b]{Jun-Bao Wu}

\affiliation[a]{State Key Laboratory of Theoretical Physics,
Institute of Theoretical Physics, Chinese Academy of Science,
Beijing 100190, P. R. China } \affiliation[b]{Institute of High
Energy Physics, and Theoretical Physics Center for Science
Facilities, Chinese Academy of Sciences, Beijing 100049, P.R. China}
\emailAdd{hesong@itp.ac.cn}\emailAdd{lidn@ihep.ac.cn}\emailAdd{wujb@ihep.ac.cn}

\date{\today}

\abstract{Potential reconstruction can be used to find various
analytical asymptotical AdS solutions in Einstein dilation system
generally. We have generated two simple solutions without physical
singularity called zero temperature solutions. We also proposed a
numerical way to obtain black hole solution in Einstein dilaton
system with special dilaton potential. By using this method, we
obtain the corresponding black hole solutions numerically and
investigate the thermal stability of the black hole by comparing the
free energy of thermal gas and the corresponding black hole. In two
groups of non-conformal gravity solutions obtained in this paper, we
find that the two thermal gas solutions are more unstable than black
hole solutions respectively. Finally, we consider black hole
solutions as a thermal state of zero temperature solutions to check
that the first thermal dynamical law exists in entanglement system
from holographic point of view.}


\begin{document}

\maketitle
\section{Introduction}
The AdS/CFT correspondence
\cite{Maldacena:1997re}\cite{Gubser:1998bc}\cite{Witten:1998qj}\cite{Aharony:1999ti}
is a very important and fundamental relation which connects
gravitational theories and quantum field theories. As an
application, in \cite{Ryu:2006bv} Ryu and Takayanagi proposed a
general way for calculating the entanglement entropy of boundary
field theory through AdS/CFT correspondence. The main point is that
the entanglement entropy in the large $N$ (and large 't Hooft
coupling) limit in field theory side can be mapped to an area of
minimal surface in gravity side. Prescription on computing the
holographic entanglement entropy (HEE) has been proved in
\cite{Lewkowycz:2013nqa}\cite{Casini:2011kv}. There are so many
evidences
\cite{Headrick:2010zt}\cite{Hartman:2013mia}\cite{Faulkner:2013yia}
to confirm this proposal  within $AdS_3/CFT_2$ correspondence. As
applications of HEE, there are intensive studies
\cite{Nishioka:2009un}-\cite{Cai:2012es} recently. More recently, in
\cite{Nozaki:2013wia}, a free falling particle in an AdS space was
used to mimic the holographic dual of local quenches and the HEE has
been computed to show the evolution of quantum entanglement. In
\cite{Hartman:2013qma}, an analytical framework for holographic
counterpart of global quantum quenches was given. In
\cite{Nozaki:2013vta}, the authors studied how a small perturbation
of HEE evolves dynamically through solving the Einstein equation in
AdS spaces.

 In vacuum state the leading divergent term of  entanglement entropy(EE) is proportional to
the area of the entangling surface (in many models)\cite{MBH}\cite{Srednicki:1993im}
 , which is the original motivation for relating EE with black hole entropy. The EE is
also an useful quantity to describe the quantum correlations between
the in and out side of a subsystem in QFT. The behavior of EE in low
excited states is also important to understand the quantum
entanglement nature of the system. This topic has been studied by
many authors, for example \cite{FMG}\cite{Masanes:2009tg}. The
elegant method of HEE could also be used to study the property of EE
in low excited states of CFT, which may be related with the
background perturbation of the bulk.

In \cite{Bhattacharya:2012mi}, the authors have studied the low
thermal excited state in the holographic view, and furthermore, they
find an interesting relation between the variance of energy and EE
of the subsystem in low thermal excited states of CFT living on the
boundary, which is similar to the first law of thermodynamics, i.e.,
$\Delta E =T_{eff} \Delta S$, where $T_{eff}$, called entanglement
temperature, is only related to the shape of the subsystem.

In effective theory of gravity, higher derivative terms will appears
as corrections to Einstein-Hilbert action. The HEE formula for
Lovelock gravity have been studied in
\cite{deBoer:2011wk}\cite{Hung:2011xb} by comparing the logarithm
term with the CFT prediction\footnote{HEE in this case was also
studied in \cite{Chen:2013qma}\cite{Bhattacharyya:2013jma} ,
following the approach of \cite{Lewkowycz:2013nqa}.}.
\cite{Ogawa:2011fw}\cite{Myers:2010xs} have also studied the HEE
with higher derivative gravity. In \cite{Guo:2013aca}, the authors
have studied the property of EE with low excitation in these cases
from the the holographical point of view. Though general formula of
HEE with the bulk theory containing arbitrary higher curvature terms
is still an open question to be further studied, one can still hope
that the results  in
\cite{deBoer:2011wk}\cite{Hung:2011xb}\cite{Chen:2013qma}\cite{Bhattacharyya:2013jma}\cite{Ogawa:2011fw}\cite{Myers:2010xs}\cite{Fursaev:2006ih}\cite{Guo:2013aca}
will shed light on the quantum corrections to HEE. More recently,
some quantum corrections have been studied in
\cite{Barrella:2013wja}\cite{Faulkner:2013ana}.

We would like to extend these studies to non-conformal cases,
especially for Einstein-dilaton system. The key point of this
extension is to find vacuum state and corresponding thermal
excitation state. Previously, there were various studies in
\cite{Gubser-T} \cite{Gursoy-T} on gravity solutions in ED system.
It is hard to obtain the gravity duals of these two states in these
frameworks. Different from the logic of \cite{Gubser-T}
\cite{Gursoy-T}, a bottom-up approach known as the potential
reconstruction approach \cite{Farakos:2009fx}\cite{Li:2011hp} is
indeed a much easier way to obtain gravity solutions. Using this
bottom-up approach, a new Schwarzschild-AdS black hole in
five-dimensions coupled to a scalar field was discussed in
\cite{Farakos:2009fx}, while dilatonic black hole solutions with a
Gauss-Bonnet term in various dimensions were discussed in
\cite{Ohta:2009pe}. A new class of four dimensional gravity
solutions has been found in \cite{Kolyvaris:2009pc}.

We will review the potential reconstruction approach to obtain
general  gravity solution in 5D. In \cite{Li:2011hp}, the authors
have used this method to construct a semianalytical gravity solution
to study some thermodynamical quantities, their results agree with
the numerical results from recent studies in lattice QCD.
\cite{Li:2011hp} provided an excellent example of constructing a
holographic model using the potential reconstruction approach.
Motivated from finding the vacuum state and thermal excitation
state, we would like to construct two gravity solutions or phases in
the same ED system. It is valuable to study the thermal excitation
properties of this system. In this paper, we will list two
analytical zero temperature solutions to show the details of this
approach.

In this paper, we would like try to obtain two black hole solutions
which correspond to these two zero temperature solutions,
respectively. To find the gravity solutions in Einstein dilation
system with special dilation potential is the hard core. We here
propose a systematical way to obtain the numerical gravity
solutions. To give the details, we take two black hole solutions
obtained in this paper as examples. Furthermore, we have studied the
free energy of these solutions and find that the thermal gas
solutions are thermal dynamically unstable. By following the logic
line proposed by \cite{Bhattacharya:2012mi}, we consider black hole
solution as the thermal excitation of corresponding zero temperature
solution and we would like study a  novel quantity called
entanglement temperature. With this entanglement temperature, there
exists the first-law-like relation that are proposed in
\cite{Bhattacharya:2012mi} then.

The organization of the paper is as follows: in section 2, we
briefly review the potential reconstruction approach to the
Einstein-Maxwell-Dilaton system by generalizing the discussion in
\cite{Cai:2012xh} to the case with a coupling between dilaton field
and Maxwell field. We also follow this approach to generate domain
wall solutions. In section 3, we discuss the generic black hole
solution with asymptotical AdS boundary in Einstein dilaton system,
and in particular present two new analytic zero temperature
solutions which will be used later. In this section, we also
proposed a numerical way to obtain the black hole solutions in ED
system. We list two groups of gravity solutions. In each group, the
one is zero temperature solution and the other is corresponding
black hole solution. In section 4, we calculate the difference of
free energy of thermal gas and the corresponding black hole
solutions generally. Here these thermal gas solutions are obtained
from Wick rotation in time direction in zero temperature solutions.
We take two groups of thermal gas and Euclidean version of black
hole solutions as examples to show  that the thermal gas solutions
are unstable. In section 5, as an application of these solutions
from AdS/CFT point of view, we study the novel quantity called
entanglement temperature in these cases and we consistently check
that the thermodynamical first law like exists in our cases. Section
6 is devoted to conclusions and discussions. We put some details of
the computations in this paper in the Appendix A. In appendix B, we
list 6 new gravity solutions generated by potential reconstruction
in ED system.

\section{Einstein-Maxwell-Dilaton system}

\label{gravitysetup}
 In this section, we just review how to  use the potential
reconstruction approach
\cite{Li:2011hp,He:2010ye,He:2011hw,Cai:2012xh} to obtain solutions
to  a 5D Einstein-Dilaton (ED) system and Einstein-Maxwell-Dilaton
(EMD) system. In \cite{Cai:2012xh}, the authors have not considered
the coupling between gauge field and dilaton field in Einstein
frame. Here we take the coupling into consideration to start with a
more general version
\begin{equation} \label{minimal-String-action}
S_{5D}=\frac{1}{16 \pi G_5}\int d^5 x \sqrt{-g^S} e^{-2 \phi}
 \left(R^S + 4\partial_\mu \phi
\partial^\mu \phi-
V_S(\phi)-\frac{Z(\phi)}{4g_{g}^2}e^{\frac{-4\phi}{3}}F_{\mu\nu}F^{\mu\nu}\right),
\end{equation}
where the action (\ref{minimal-String-action}) is written in string
frame, $F_{\mu\nu}=\partial_\mu A_\nu-\partial_\nu A_\mu$ is the
Maxwell field. In Einstein frame, we can write the action as
\cite{He:2010ye}
\begin{equation} \label{minimal-Einstein-action}
S_{5D}=\frac{1}{16 \pi G_5} \int d^5 x
\sqrt{-g^E}\left(R-\frac{4}{3}\partial_{\mu}\phi\partial^{\mu}\phi-V_E(\phi)
-\frac{Z(\phi)}{4g_{g}^2}F_{\mu\nu}F^{\mu\nu}\right),
\end{equation}
where $ V_S=V_E e^{\frac{-4\phi}{3}}.$ The metrics in both frames
are connected by the scaling transformation
\begin{equation}
g^S_{\mu\nu} = e^{4 \phi \over 3 }g^E_{\mu\nu}.
\end{equation}
The Einstein equations from the action
(\ref{minimal-Einstein-action}) read
\begin{eqnarray} \label{EOM}
E_{\mu\nu}+\frac{1}{2}g^E_{\mu\nu}\left(\frac{4}{3}
\partial_{\mu}\phi\partial^{\mu}\phi+V_E(\phi)\right)
-\frac{4}{3}\partial_{\mu}\phi\partial_{\nu}\phi -\frac{Z(\phi)}{2
g_{g}^2}\left(F_{\mu k}{F_\nu}^k-\frac{1}{4}
g^E_{\mu\nu}F_{kl}F^{kl}\right)=0,
\end{eqnarray}
where $E_{\mu\nu}=R_{\mu\nu}-\frac{1}{2}Rg_{\mu\nu}$ is Einstein
tensor. When we turn off the gauge field, the EDM system will be
reduce to ED system given in appendix A. We here consider the ansatz
$A=A_0(z) dt,\text{{ }} \text{{ }}\phi=\phi(z)$ for matter fields
and
\begin{equation} \label{metric-stringframe}
ds_S^2 = \frac{{L^2} e^{2A_s}}{z^2}
\left(-f(z)dt^2+\frac{dz^2}{f(z)}+dx^{i}dx^{i}\right),
\end{equation}
for the metric in string frame, where $L$ is the radius of ${\rm
AdS}_5$ space and $A_s$ is the warped factor, a function of
coordinate $z$. 
In Einstein frame the metric
reads
\begin{eqnarray} \label{metric-Einsteinframe}
ds_E^2&= &\frac{{L^2} e^{2A_e}}{z^2}\left(-f(z)dt^2
+\frac{dz^2}{f(z)}+dx^{i}dx^{i}\right),\nonumber\\
&=& \frac{{L^2} e^{2A_s-\frac{4\phi}{3}}}{z^2}\left(-f(z)dt^2
+\frac{dz^2}{f(z)}+dx^{i}dx^{i}\right),
\end{eqnarray}
with $A_e=A_s-2\phi/3$.

In the metric (\ref{metric-Einsteinframe}), the $(t,t), (z,z)$ and
$(x_1, x_1)$ components of Einstein equations are respectively
\bea\label{Einsteiineq} &{}&b''(z)+\frac{b'(z) f'(z)}{2
f(z)}-\frac{b'(z)^2}{2 b(z)}+\frac{4}{9} b(z) \phi
'(z)^2+\frac{{A_0}'(z)^2 Z(\phi)}{6 g^2_g f(z)}+\frac{V_E(\phi)
b(z)^2}{3 f(z)}=0, \nonumber\\
&{}& \phi '(z)^2-\frac{9 b'(z) f'(z)}{8 b(z) f(z)}-\frac{9
b'(z)^2}{4 b(z)^2}-\frac{3 {A_0}'(z)^2 Z(\phi)}{8 g^2_g b(z)
f(z)}-\frac{3 V_E(\phi) b(z)}{4 f(z)}=0,
\nonumber\\
&{}&f''(z)+\frac{3 b'(z) f'(z)}{b(z)}+\frac{4}{3} f(z) \phi
'(z)^2+\frac{3 f(z) b''(z)}{b(z)}-\frac{3 f(z) b'(z)^2}{2
b(z)^2}-\frac{{A_0}'(z)^2 Z(\phi)}{2 g^2_g b(z)}+V_E(\phi) b(z)=0
,\nonumber \\
\eea
 where $b(z)=\frac{L^2 e^{2A_e}}{z^2} $, and
$A_0(z)$ is electrical potential of Maxwell field. From those three
equations one can obtain following two equations which do not
contain the dilaton potential $V_E(\phi)$,
\begin{eqnarray}\label{AF}
 &{}&A_s''(z)+A_s'(z) \left(\frac{4 \phi '(z)}{3}+\frac{2}{z}\right)-A_s'(z)^2-\frac{2 \phi ''(z)}{3}-\frac{4 \phi '(z)}{3 z}=0,\\\label{ff}
 &{}&f''(z)+ f'(z)\left(3 A_s'(z)-2 \phi '(z)-\frac{3 }{z}\right)-\frac{z^2 Z(\phi)e^{\frac{4 \phi (z)}{3}-2 A_s(z)} A_0'(z){}^2}{ g_{g}^2 L^2}=0.
\end{eqnarray}
Eq.(\ref{AF}) is the starting point to find exact solutions of the
system.  Note that Eq.(\ref{AF}) in the EMD system is the same as
the one in the Einstein-dilaton system considered in
\cite{Li:2011hp}\cite{He:2011hw} and the last term in Eq.(\ref{ff})
is an additional contribution related to electrical field. In
addition, the EOM of the dilaton field is given as following
\begin{equation}
\label{fundilaton} \frac{8}{3} \partial_z
\left(\frac{L^3e^{3A_s(z)-2\phi} f(z)}{z^3}
\partial_z \phi\right)-
\frac{L^5e^{5A_s(z)-\frac{10}{3}\phi}}{z^5}\partial_\phi V_E(\phi)+
\frac{Z'(\phi)b(z) A_0'(z)^2}{2 g_g^2}=0.
\end{equation}
The EOM of the Maxwell field is given as \bea \frac{1}{\sqrt{-g^E}}
\partial_\mu \left(\sqrt{-g^E}Z(\phi) F^{\mu\nu}\right)=0.\eea

From equations of motion, we can obtain a general solution to the
system with given $A_s(z)$, which takes the following form \bea
\label{solutionU(1)1}\phi(z)&=&\int_0^z \frac{e^{2A_s(x)}
\left(\frac{3}{2} \int_0^x y^2 e^{-2 A_s(y)} A_s'(y)^2 \, dy+\phi
_1\right)}{x^2} \, dx+\frac{3 A_s(z)}{2}+\phi _0,\\
A_0(z)&=&A_{00}+A_{01} \left(\int_0^z {y e^{\frac{2 \phi
(y)}{3}-A_s(y)} \over Z(\phi(y))}\, dy\right),\\ \label{f(z)}
f(z)&=&\int _0^zx^3 e^{2 \phi (x)-3 A_s(x)} \left(\frac{A_{01}{}^2
\left(\int_0^x  {y e^{\frac{2 \phi
(y)}{3}-A_s(y)}\over Z(\phi(y))}\, dy\right)}{ g_{g}^2 L^2}+f_1\right)dx+f_0,\\
V_E(z)&=&\frac{e^{-2 A_s(z)+\frac{4 \phi(z)}{3}} z^2 f(z)}{L^2}2
\Big(-\frac{e^{-2 A_s(z)+\frac{4 \phi(z)}{3}}Z(\phi(z)) z^2
A_0'(z)^2}{4 g_g^2L^2 f(z)}\nonumber\\&-&\frac{2 \left(3+3 z^2
A_s'(z)^2+4 z \phi'(z)+z^2 \phi'(z)^2-2 zA_s'(z) \left(3+2 z
\phi'(z)\right)\right)}{z^2}\nonumber\\&-&\frac{f'(z) \left(-3+3 z
A_s'(z)-2 z\phi'(z)\right)}{2 z f(z)}\Big),\label{solutionU(1)4}
 \eea
where the $\phi_0, A_{00}, A_{01}, f_0, f_1$ are all integration
constants and can be determined by suitable UV and IR boundary
conditions. Specially for $Z(\phi)=1$, the general solution reduces
to the one given in \cite{Cai:2012xh}. Thus we have reviewed a
generic formalism \cite{Cai:2012eh} to generate exact solutions of
the EMD system with a given $A_s(z)$.

\subsection{Potential reconstruction in domain wall ansatz}
By using of potential reconstruction approach, one can  generate
gravity solution as he/she wants. In this subsection, we do not
repeat the details and  just list the results.

Our starting point is the following Einstein-Maxwell-dilaton action in $d+1$
spacetime dimensions:

\begin{eqnarray}
&&I={1\over 16\pi G_{d+1}}\int_{{\cal M}}
d^{d+1}x\sqrt{-g}[R-\frac{Z(\phi)}{4}F^2-\frac{1}{2}(\partial\phi)^2+V(\phi)]+I_{GH},\\
&&I_{\text{GH}}={2\over 16\pi G_{d+1}}\int_{\partial{\cal
M}}d^{d}x\sqrt{h}K.
\end{eqnarray}
This action describes the dynamics of a $U(1)$ gauge field ${\bf
A}_\mu$ (with field strength $F=d{\bf A}$) and a real scalar field
$\phi$ coupled to Einstein gravity. The boundary term
$I_{\text{GH}}$ is the standard Gibbons-Hawking term needed to make
the variational problem well-defined.
As such this action describes the grand
canonical ensemble.

Since we are interested in solutions with finite temperature and
chemical potential and we set following domain wall ansatz
\begin{eqnarray}\label{domainwallansatz}
&& ds^2=e^{2A(u)}(-f(u)dt^2+dx^{i}dx^{i})+\frac{du^2}{f(u)}, {\bf
A}=A_{t}(u)dt, \phi=\phi(u),
\end{eqnarray}
where the AdS radius has been set to one. In this frame the second
order equations of motion reduce to the following set of
differential equations

\begin{eqnarray}
&&\frac{d}{du}(e^{(d-2)A}Z\dot{A}_{t})=0,\\
&&2(d-1)\ddot{A}+\dot{\phi}^2=0,\\
&&\ddot{f}+d\dot{A}\dot{f}-e^{-2A}Z\dot{A}^2_{t}=0,\\
&&(d-1)\dot{A}\dot{f}+(d(d-1)\dot{A}^2-\frac{1}{2}\dot{\phi}^2)f-V+\frac{1}{2}Ze^{-2A}\dot{A}^2_{t}=0.
\end{eqnarray} Here the dot stands for derivative with respect to
$u$. We can extend the logic of potential reconstruction to this
general case with domain wall ansatz. The general solutions are as
follows
 \bea\label{solutiondomainwall}
A_{t}(u)&=&\int_1^u \frac{\rho e^{-2 (A( x))}}{Z (\phi( x))} \,
dx+A_{t0},\nonumber\\ A &=& a_1 u+\int_0^u \left(\int_0^y
-\frac{1}{6} \phi'(x)^2 \, dx\right) \,
dy,\nonumber\\f(u)&=&\left.\int_0^u \left(e^{-4 (A y)} \int_0^y e^{2
(A x)} \left( A_{t}(x)\right)^2 (Z (\phi(x))) \, dx+f_1
e^{-4 (A y)}\right) \, dy+f_0\right),\nonumber\\
V(u)&=&(d-1) \left( A'(u)\right) \left(f'(u)\right)+(f( u))
\left((d-1) d \left( A'(u)\right)^2-\frac{1}{2} \left(
\phi'(u)\right)^2\right)\nonumber\\&+&\frac{1}{2} e^{-2 (A( u))}
\left( A_{t}'(u)\right)^2 (Z (\phi(u))), \eea where $A_{t0}, a_1,
f_0, f_1 $ are integral constants. One can use the $A(u)$ defined in
(\ref{domainwallansatz}) to generate the whole gravity solutions in
terms of ({\ref{solutiondomainwall}}). As an application, we here
list one explicit solution with $d=4$:
\begin{eqnarray}
A(u)&=&u+a,\nonumber\\
\phi(u)&=&b,\nonumber\\
A_t(u)&=&-\frac{c}{2}e^{-2u},\nonumber\\
f(u)&=&1+\frac{c^{2}e^{-2a-6u}}{12}-\frac{de^{-4u}}{4}.\nonumber\\
V(u)&=&12.
\end{eqnarray}
Here $a, b, c, d$ are integral constants which can be determined by
boundary conditions. One can set $a=b=c=d=0$ to reproduce the
pure $AdS_5$ solution. This example is used to show this method is
convenient to generate the gravity solution effectively.

\section{General asymptotical AdS black hole solutions}
\label{general solution} In this part, we will review the general
asymptotical AdS black hole solutions given in
\cite{Cai:2012eh}. Since we are only interested in the black hole
solutions with asymptotic $AdS$ boundary, \cite{Cai:2012eh} impose
the boundary condition $f(0)=1$ at the AdS boundary  $z= 0$, and
require $\phi(z), f(z), A_0(z)$ to be regular at black hole horizon
$z_h$ and AdS boundary $z=0$. There is an additional condition
$A_0(z_h)=0$, which corresponds to the physical requirement that
$A_\mu A^\mu=g^{tt}A_0A_0$ must be finite at $z=z_h$.

\cite{Cai:2012eh} expressed the function $f(z)$ in
eq.(\ref{f(z)}) as
 \bea\label{ffunction}
  f(z)&=&1+
\frac{A_{01}{}^2 }{2 g_{g}^2 {L^2}}\frac{\int_0^z
g(x)\left(\int_0^{z_h}g(r)dr \int_r^x {g(y)^{\frac{1}{3}}dy\over
Z(\phi(y))}\right)dx}{\int_0^{z_h}g(x)dx} -\frac{\int_0^z g(x)dx
}{\int_0^{z_h}g(x)dx},
 \eea
 where $f_0=1$,
$f_1=-\frac{A_{01}{}^2 }{4g_{g}^2
{L^2}}\frac{\int_0^{z_h}g(x)\int_0^x{g(y)^{\frac{1}{3}}\over
Z(\phi(y))}dy+1}{\int_0^{z_h}g(x)dx}$ and
 \bea \label{xyfunction}
g(x)&=&x^3 e^{2 \phi (x)-3 A_s(x)}. \eea One can expand the gauge
field near the AdS boundary to relate the two integration constants
to chemical potential and charge density \bea A_0(z)&\sim &
A_{00}+A_{01}{e^{\frac{2\phi(y)}{3}-A_s(y)}\over
Z(\phi(y))}\left(1+y(\frac{2\phi'(y)}{3}-A_s'(y))\right)\Big|_{y=0}
z^2,\label{chemical} \eea with
 \bea A_{00} &=&\mu,\\
A_{01}&=& \frac{\mu}{\int_0^{z_h}y {e^{\frac{2\phi}{3}-A_s(y)}\over
Z(\phi(y))}dy}=\frac{\mu}{\int_0^{z_h} {g(y)^{\frac{1}{3}}\over
Z(\phi(y))}dy}. \eea

The temperature of the black hole can be determined through the
function $f(z)$ in (\ref{ffunction}) as   \bea\label{temp}
T=\left|{A^2_{01}\over 4\pi g_g^2 L^2}\frac{
g(z_h)\int_0^{z_h}g(r)dr\int_r^{z_h}{g^{1\over 3}(y)\over
Z(\phi(y))}dy-g(z_h)}{\int_0^{z_h}g(x)dx}\right|. \eea Following the
standard Bekenstein-Hawking entropy formula, from the geometry given
in eq.~(\ref{metric-Einsteinframe}), we obtain the black hole entropy
density $S$, which is obtained using the area $A_{area}$ of the horizon
\begin{equation}
\label{entrpy} S={\frac{A_{area}}{4 G_5 V_3}=
\frac{L^3}{4G_5}\left(\frac{e^{A_s-\frac{2}{3}\phi}}{z}\right)^3}\Big|_{z_h},
\end{equation}
where $V_3$ is the volume of the black hole spatial directions
spanned by coordinates $x_i$ in (\ref{metric-Einsteinframe}). For
this paper, we do not consider the effects of the $U(1)$ gauge
field. In the remain part, we will focus on the ED system. We will
propose an algorithm to obtain the black hole solution in ED system
numerically. The algorithm is different from the potential
construction approach shown in eq. (\ref{solutionU(1)1})-eq.
(\ref{solutionU(1)4}). Here we just review the previous results and
obtain two zero temperature solutions in different ED systems. It is
just technical trick to obtain zero temperature solutions. In the
following subsection, we will focus on how to obtain the
corresponding black hole solution numerically with respect to
special zero temperature solutions in ED system.

\subsection{The first analytical solution}
\label{appendix-solution} In this subsection, we list an analytical
 solution of the Einstein-Maxwell-Dilaton system by using
Eq.(\ref{solutionU(1)1}-\ref{solutionU(1)4}) with $Z(\phi)=1$. We
impose the constrain  $f(0)=1$, and require $\phi(z), f(z)$ to be
regular at $z=0$, and $z_h$. We give the solution in Einstein frame
\begin{equation} \label{Ametric-Einsteinframe}
ds_E^2 =\frac{{L^2} e^{2A_{e1}}}{z^2}\left(-f_1(z)dt^2
+\frac{dz^2}{f_1(z)}+dx^{i}dx^{i}\right),
\end{equation}
with
\begin{eqnarray}
A_{e1}(z)&=&\log \left(\frac{z }{z_0\sinh(\frac{z}{z_0})}\right),\nonumber\\
f_1(z)&=&1-\frac{4 V_{11}}{3}(3\sinh(\frac{z}{z_0})^4+2\sinh(\frac{z}{z_0})^6)+\frac{1}{8} V_{12}^2 \sinh\left(\frac{z}{z_0}\right)^4,\nonumber\\
\phi_1(z)&=&\frac{3 z}{2 z_0},\nonumber\\
A_{01}(z)&=& \mu_1 -\frac{2g_g L}{z_0} V_{12}
\sinh\left(\frac{z}{2z_0}\right)^2, \label{sol1}
\end{eqnarray}
where $z_0$ is an integration constant and $V_{11},V_{12}$ are two
constants from the dilaton potential \bea\label{dilatonpotential1}
V_{E1}(\phi_1)&=&-\frac{12+9\sinh^2\left(\frac{2\phi_1}{3}\right)
+16V_{11}\sinh^6\left(\frac{\phi_1}{3}\right)}{{L^2}}+\frac{V_{12}^2
\sinh^6\left(\frac{2 \phi_1 }{3}\right)}{8 {L^2}}. \eea Note that
$V_{11}$ can be parameterized by  black hole horizon $z_h$. The two
integration constants $V_{11}$ and $V_{12}$ then can be expressed in
terms of horizon $z_h$ as  \bea
V_{11}&=&\frac{3{\cosh}^4\left(\frac{z_h}{2 z_0}\right)
\left(\frac{\mu ^2 z_0^2 \sinh ^4\left(\frac{z_h}{z_0}\right)
{\cosh}^4\left(\frac{z_h}{2 z_0}\right)}{4 g_g^2 {L^2}}+8\right)}{32
\left(2 \sinh ^2\left(\frac{z_h}{2
z_0}\right)+3\right)},\nonumber\\V_{12}&=& \frac{\mu z_0
{\cosh}^2\left(\frac{z_h}{2 z_0}\right)}{2 g_g L}. \eea

In this solution, we can just only consider the degenerate case with
$V_{11}=0, V_{12}=0, \mu_1=0$. The explicit form for the degenerate
case called the first zero temperature solution is 
\begin{eqnarray}\label{firstsolution}
A_{et1}(z)&=&\log \left(\frac{3 p_1z }{2\sinh(\frac{3 p_1z}{2})}\right),\nonumber\\
f_{t1}(z)&=&1,\nonumber\\
\phi_{t1}(z)&=& p_1 z,\nonumber\\
 V_{Et1}(\phi)&=&-\frac{12+9\sinh^2\left(\frac{2\phi_{t1}}{3}\right)
}{{L^2}}. \eea In the last step we have set $z_0= {2\over 3}p_1$
which is helpful for following analysis.

\subsubsection{The corresponding black hole solution}
In this subsection, we would like to find the black hole solution
with the same potential as the previous subsection in ED system.
Firstly, the UV behavior of the black hole should be asymptotical
AdS and there is a horizon in the IR region which is parameterized
by $z_h$. We find an algorithm to find the numerical solution
consistently. In order to show how the algorithm work, we list all
the details in appendix. Roughly speaking, we try to  expand in
power series all unknown function as positive powers of $z$ and try
to fix all the coefficients numerically. In the UV region, the black
hole can be solved as follows from coupled equations of motion with
the potential Eq.~(\ref{firstsolution})\bea
\label{firstblackhole}\phi_{b1}(z)&=&p_1 z + p_3 z^3 + {((405 f_{41}
p_1 +
    612 p_1^2 p_3) z^5)\over 3240 }\nonumber\\&+& {(8100 f_{41} p1^3 + 229635 f_{41} p_3
    +
    10944 p_1^4 p_3 + 133164 p_1 p_3^2) z^7\over 612360}+O(z^7)\nonumber\\
    f_{b1}(z)&=&1 - f_{41} z^4 -
{4\over 27} f_{41} p_1^2 z^6 + {-13 f_{41} p_1^4\over 1215} -
{f_{41} p_1
 p_3\over
    5}) z^8 \nonumber\\&+& {(-10935 f_{41}^2 p_1^2 - 328 f_{41} p_1^6 - 37908 f_{41} p_1^3 p_3 -
    78732 f_{41} p_3^2) z^{10}\over 688905}+O(z^{10})\nonumber\\
   A_{eb1}(z)&=& -(2/27) p_1^2 z^2 + ({(4 p_1^4)\over 3645} - {(2 p_1
   p_3)\over
    15}) z^4 \nonumber\\&+& {(-54675 f_{41} p_1^2 - 128 p_1^6 - 67068 p_1^3 p_3 -
    393660 p_3^2) z^6)\over 4133430} \nonumber\\&+& {((-50625 f_{41} p_1^4 + 64 p_1^8 -
    3444525 f_{41} p_1 p_3 - 74952 p_1^5 p_3 -
    2943216 p_1^2 p_3^2) z^8\over 62001450}\nonumber\\&+&O(z^{8}).\eea
One can see the black hole solution in the UV region can be
expressed in series of powers of $z$. In principle, one can obtain
more higher powers of $z$ to get the full expression of black hole
background. Unfortunately, we can not obtain complete form of the
black hole solution. The main reason is that we do not find simple
recurrence relation among the coefficients of each power of $z$. It
is easy to see that the black hole solution with asymptotical AdS
can be controlled by three integral constants $p_1,p_3, f_{41}$.
$p_1$ is free parameter and $p_3, f_{41}$ are determined by boundary
condition in IR region. Here we choose parameters $p_1, p_3, f_{41}$
to show one black hole solution numerically. Here $ p_3, f_{41}$ are
not independent and they are related to the horizon position $z_h$
such that $f_{b1}(z_h)=0$. In appendix A, we will show the details
how to find $z_h$. The numerical relation between $ p_3(z_h)$ and $
f_{41}(z_h)$ has been shown in Fig.~\ref{f4p3-3}.

\begin{figure}[h]
\begin{center}
\epsfxsize=7.5 cm \epsfysize=5.5 cm \epsfbox{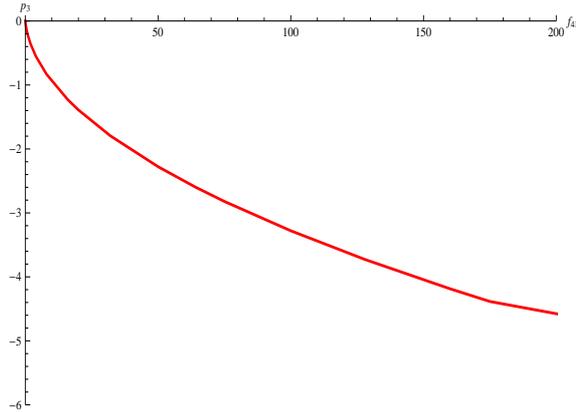}
\end{center}
\caption{$p_3 \text{v.s.} f_4$.  $p_{3} $ decreasing with $f_{41}$
monotonously. Here we have chosen the parameter $p_{1}=
1.5\text{GeV}$.}\label{f4p3-3}
\end{figure}

For simplifying numerical analysis, we fix $p_1=1.5 \text{GeV}$.
Here we just only show the series expansion of black hole solutions
and the complete solution can be obtained by the algorithm explained
in appendix A. Figs.~\ref{phi-3}-\ref{Ae-3} show the configuration
of black hole solution. The black hole horizon can be easily read
out by $f_{b1}(z_h)=0$. From these figures, one can continuously go
back to zero temperature solution from black hole solution with
decreasing value of $f_{41}$ to zero. $f_{41}$ decreases with
increasing $p_3$ simultaneously. Where the temperature is defined by
$T={{f_{b1}(z)'\over 4\pi}|_{z=z_h}}$. This behavior will be helpful
to understand entanglement temperature in section 5.

\begin{figure}[h]
\centering
\includegraphics[width=8 cm]{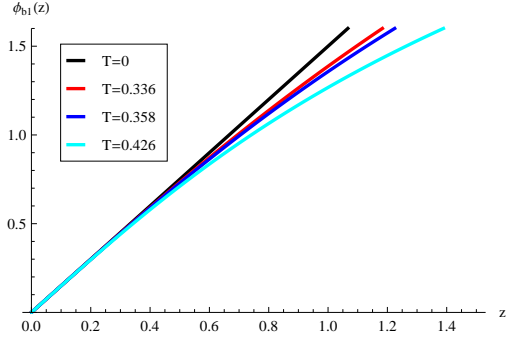}
\caption{The configuration of $\phi_{b1}(z)$ as a function
holographic coordinate $z$. Here we have fixed $p_1=1.5 \text{GeV}$
all over the paper. The four parameters for these curves are taken
values as following \{${T=0, f_{41}=0, p_3=0}$\}, \{${T=0.336
\text{GeV}, f_{41}=0.5 \text{GeV}^4, p_3=-0.139... \text{GeV}^3}$\},
\{${T=0.358 \text{GeV}, f_{41}=0.75 \text{GeV}^4, p_3=-0.186...
\text{GeV}^3}$\}, \{${T=0.426,f_{41}=2\text{GeV}^4, p_3=-0.360...
\text{GeV}^3}$\} respectively. To get a smooth function numerically,
$p_3$ have very high-precision and we replace the more additional
number with "..." for short.} \label{phi-3}
\end{figure}

\begin{figure}[h]
\centering
\includegraphics[width=8 cm]{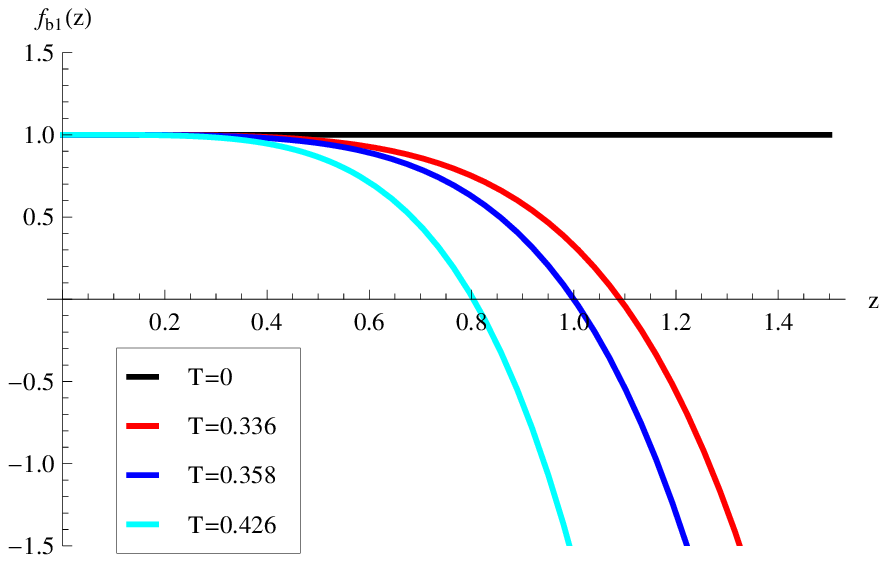}
\caption{The configuration of $f_{b1}(z)$ as a function holographic
coordinate $z$. Here we have fixed $p_1=1.5 \text{GeV}$ all over
the paper. The four parameters for these curves are taken values as
following \{${T=0, f_{41}=0, p_3=0}$\}, \{${T=0.336 \text{GeV},
f_{41}=0.5 \text{GeV}^4, p_3=-0.139... \text{GeV}^3}$\}, \{${T=0.358
\text{GeV}, f_{41}=0.75 \text{GeV}^4, p_3=-0.186...
\text{GeV}^3}$\}, \{${T=0.426,f_{41}=2\text{GeV}^4, p_3=-0.360...
\text{GeV}^3}$\} respectively. To get a smooth function numerically,
$p_3$ have very high-precision and we replace the more additional
number with "..." for short.}\label{ff-3}
\end{figure}

\begin{figure}[h]
\centering
\includegraphics[width=8 cm]{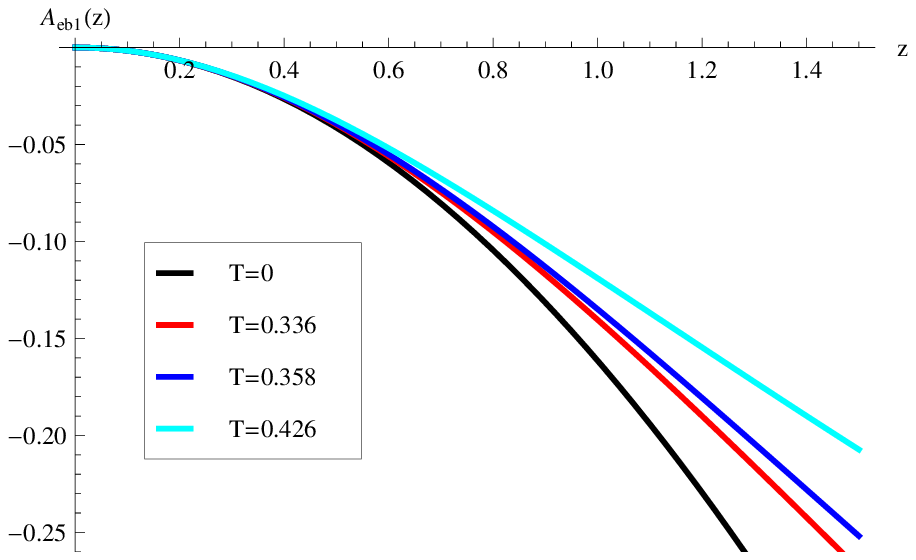}
\caption{The configuration of $A_{eb1}(z)$ as a function holographic
coordinate $z$. Here we have fixed $p_1=1.5 \text{GeV}$ all over
the paper. The four parameters for these curves are taken values as
following  \{${T=0, f_{41}=0, p_3=0}$\}, \{${T=0.336 \text{GeV},
f_{41}=0.5 \text{GeV}^4, p_3=-0.139... \text{GeV}^3}$\}, \{${T=0.358
\text{GeV}, f_{41}=0.75 \text{GeV}^4, p_3=-0.186...
\text{GeV}^3}$\}, \{${T=0.426,f_{41}=2\text{GeV}^4, p_3=-0.360...
\text{GeV}^3}$\} respectively. To get a smooth function numerically,
$p_3$ have very high-precision and we replace the more additional
number with "..." for short.} \label{Ae-3}
\end{figure}

\subsection{The second analytical solution}

The second  exact solution with ansatz \begin{equation}
\label{Ametric-Einsteinframe} ds_E^2 =\frac{{L^2}
e^{2A_{e2}}}{z^2}\left(-f_2(z)dt^2 +\frac{dz^2}{f
_2(z)}+dx^{i}dx^{i}\right),
\end{equation} is \bea
A_{e2}(z)&=&-\log \left(1+\frac{z}{z_0}\right),\\
f_2(z)&=& 1-V_{21} \left(\frac{z^7}{7 z_0^7}+\frac{z^6}{2 z_0^6}+
+\frac{3 z^5}{5 z_0^5}+\frac{z^4}{4 z_0^4}\right)\nonumber\\
&&+\frac{\rho_2^2 z_0^8}{ g_g^2 L^2}\Big( \frac{5 z^8 }{32
z_0^8}+\frac{z^{10} }{60
 z_0^{10}} +\frac{z^9}{12 z_0^9}+\frac{11 z^7 }{84 z_0^7}
+\frac{z^6 }{24z_0^6}\Big),\\
\phi_2(z)&=&3 \sqrt{2} \sinh
^{-1}\left(\sqrt{\frac{z}{z_0}}\right),\nonumber\\
A_{02}(z)&=& \mu_2+\rho_2  \left(\frac{z_0
z^2}{2}+\frac{z^3}{3}\right).\eea where $z_0,\mu_2,$ and  $\rho_2$
are integration constants and $V_{21}$ is a constant from the
dilaton potential and $g_g$ is gauge coupling.  The dilaton
potential is given as \bea\label{dilatonpotential2}
V_{E2}(\phi_2)&=&-\frac{12}{L^2}-\frac{42 \sinh ^4\left(\frac{\phi_2
}{3 \sqrt{2}}\right)}{L^2}-\frac{42 \sinh ^2\left(\frac{\phi_2 }{3
\sqrt{2}}\right)}{L^2}\nonumber\\&{}&-\frac{3 V_{21} \sinh
^{14}\left(\frac{\phi_2 }{3 \sqrt{2}}\right)}{35 L^2}-\frac{3 V_{21}
\sinh ^{12}\left(\frac{\phi_2 }{3 \sqrt{2}}\right)}{10
L^2}-\frac{3V_{21} \sinh ^{10}\left(\frac{\phi_2 }{3
\sqrt{2}}\right)}{10
L^2}\nonumber\\
&{}&+\frac{\rho^2 z_0^8}{ g_g^2 L^2}\Big\{\frac{ \sinh
^{24}\left(\frac{\phi_2 }{3 \sqrt{2}}\right)}{20  L^2}+\frac{3 \sinh
^{22}\left(\frac{\phi_2 }{3 \sqrt{2}}\right)}{10 L^2}+\frac{59 \sinh
^{20}\left(\frac{\phi_2 }{3
\sqrt{2}}\right)}{80L^2}\nonumber\\&{}&+\frac{15  \sinh
^{18}\left(\frac{\phi_2 }{3 \sqrt{2}}\right)}{16 L^2}+\frac{5 \sinh
^{16}\left(\frac{\phi_2 }{3 \sqrt{2}}\right)}{8 L^2}+\frac{5 \sinh
^{14}\left(\frac{\phi_2 }{3 \sqrt{2}}\right)}{28 L^2}\Big\}. \eea
This solution is also a generalization of the one given in
\cite{Li:2011hp}.

Here we turn off the $U(1)$ gauge field and the second solution can
be reduced to the following \bea\label{secondsolution}
A_{et2}(z)&=&-\log \left(1+p_{1\over 2}^2 z\right),\\
f_{t2}(z)&=& 1,\\
\phi_{t2}(z)&=&3 \sqrt{2} \sinh
^{-1}\left(p_{1\over 2}\sqrt{ z}\right),\nonumber\\
V_{Et2}(\phi_{t2})&=&-\frac{12}{L^2}-\frac{42 \sinh
^4\left(\frac{\phi_{t2} }{3 \sqrt{2}}\right)}{L^2}-\frac{42 \sinh
^2\left(\frac{\phi_{t2} }{3 \sqrt{2}}\right)}{L^2}.\eea Which is so
called the second zero temperature solution. In the last step we
have set $z_0= {1\over \sqrt{p_{1\over 2}}}$ which is helpful for
following analysis.
\subsubsection{The corresponding black hole solution}
In this subsection, we would like to find the black hole solution
with same potential in Einstein dilation system. As before, the UV
behavior of the black hole should be asymptotical AdS and there is a
horizon in the IR region which is parameterized by $z_h$.  The
series expansion of solution near UV region is:
\bea\label{BHsolution2} \phi_{b2}(z)&=&p_{\frac{1}{2}} \sqrt{z}
-\frac{1}{108} p_{\frac{1}{2}}^3 z^{3/2}+\frac{p_{\frac{1}{2}}^5
z^{5/2}}{4320}+\frac{\left(653184 p_{\frac{7}{2}}-5
p_{\frac{1}{2}}^7\right) z^{7/2}}{653184}
\nonumber\\&+&\frac{z^{9/2} \left(18895680 f_{42}
p_{\frac{1}{2}}+515 p_{\frac{1}{2}}^9+\frac{171}{2} \left(653184
p_{\frac{7}{2}}-5 p_{\frac{1}{2}}^7\right)
p_{\frac{1}{2}}^2\right)}{302330880}\nonumber\\&+&\frac{z^{11/2}
\left(69284160 f_{42} p_{\frac{1}{2}}^3+1135
p_{\frac{1}{2}}^{11}+\frac{517}{2} \left(653184 p_{\frac{7}{2}}-5
p_{\frac{1}{2}}^7\right)
p_{\frac{1}{2}}^4\right)}{13302558720}\nonumber\\&+& O(z^{11\over
2}),
\nonumber\\
f_{b2}(z)&=&1-f_{42} z^4-\frac{1}{15} 2 f_{42} p_{\frac{1}{2}}^2
z^5-\frac{1}{162} f_{42} p_{\frac{1}{2}}^4 z^6-\frac{f_{42}
p_{\frac{1}{2}}^6 z^7}{10206}\nonumber\\& +&z^8 \left(-\frac{f_{42}
p_{\frac{1}{2}}^8}{1119744}-\frac{f_{42} \left(653184
p_{\frac{7}{2}}-5 p_{\frac{1}{2}}^7\right)
p_{\frac{1}{2}}}{5598720}\right) \nonumber\\&+& \frac{z^9 \left(-35
f_{42} p_{\frac{1}{2}}^{10}-7 f_{42} \left(653184 p_{\frac{7}{2}}-5
p_{\frac{1}{2}}^7\right) p_{\frac{1}{2}}^3-839808 f_{42}^2
p_{\frac{1}{2}}^2\right)}{151165440}\nonumber\\& +&O(z^{10}), \eea

\bea
 A_{eb2}(z)&=&-\frac{1}{18} p_{\frac{1}{2}}^2 z+\frac{1}{648}
p_{\frac{1}{2}}^4 z^2-\frac{p_{\frac{1}{2}}^6 z^3}{17496} +
\left(\frac{p_{\frac{1}{2}}^8}{559872}-\frac{p_{\frac{1}{2}}
\left(653184 p_{\frac{7}{2}}-5
p_{\frac{1}{2}}^7\right)}{8398080}\right) z^4\nonumber\\&+&
\frac{z^5 \left(-2519424 f_{42} p_{\frac{1}{2}}^2-109
p_{\frac{1}{2}}^{10}-9 \left(653184 p_{\frac{7}{2}}-5
p_{\frac{1}{2}}^7\right)
p_{\frac{1}{2}}^3\right)}{604661760}\nonumber\\& +& \frac{z^6
\left(-37791360 f_{42} p_{\frac{1}{2}}^4+325
p_{\frac{1}{2}}^{12}-159 \left(653184 p_{\frac{7}{2}}-5
p_{\frac{1}{2}}^7\right)
p_{\frac{1}{2}}^5\right)}{228562145280}\nonumber\\&+& O(z^{6}).\eea
One can see the black hole solution in the UV region can be
expressed in series of powers of $z$. In principle, one can obtain
more higher powers of $z$ to produce the full expression of black
hole background. Unfortunately, we can not obtain complete form of
the black hole solution. The main reason is still that we also do
not find recurrence relation among the coefficients of each power of
$z$. It is easy to see that the black hole solution with
asymptotical AdS can be controlled by three integral constants
$p_{1\over 2},p_{7\over 2}, f_{42}$. $p_{1\over 2}$ is free and
$p_{1\over 2},p_{7\over 2}, f_{42}$ are determined by boundary
condition in IR region. In this case, $p_{7\over 2}, f_{42}$ are not
independent and they are determined by the black hole horizon $z_h$.
The numerical relation between $p_{7\over 2}$ and $f_{42}$ has been
shown in Fig.~\ref{f4p3-4}.

\begin{figure}[h]
\begin{center}
\epsfxsize=7.5 cm \epsfysize=5.5 cm \epsfbox{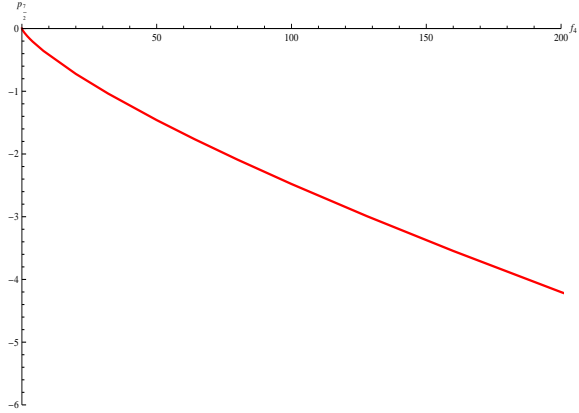}
\end{center}
\caption{$p_{7\over 2} \text{v.s.} f_{42}$. $p_{7\over 2} $
decreasing with $f_{42}$ monotonously. Here we have chosen the
parameter $p_{1\over 2}= 1\text{GeV}$.}\label{f4p3-4}
\end{figure}

For simplifying, we choose groups of $p_{1\over 2}=1
\text{GeV}^{1\over 2}$ in this paper. Here we fix parameters
$p_{7\over 2}, f_{42}$ to show black hole solutions numerically in
Fig.~\ref{phi-4}-\ref{Ae-4}. Finally, one can obtain the zero
temperature solution with setting $p_{1\over 2}=1 \text{GeV}^{1\over
2}, p_{7\over 2}=0, f_{42}=0$. Tuning on $p_{7\over 2}, f_{42}$
correspond to thermal excitation of zero temperature solution and
one also have seen this phenomenon in the first group of solution.
In the next section, we will calculate the difference of free energy
between the Euclidean black hole and thermal gas obtained by
Euclidean zero temperature solution to prove thermal gas is more
unstable than black hole. In this case, turning on small value of
$f_{42}$ correspond the thermal excitation from zero temperature
solution. Simultaneously, $f_{42}$ increases from zero to small
positive value corresponds that decreasing $p_{7\over 2}$ from zero
to small negative number. That is to say the black hole solution can
be degenerated to zero temperature solution continuously with
setting $f_{42}=0, p_{7\over 2}=0$.Where the temperature is defined
by $T={{f_{b2}(z)'\over 4\pi}|_{z=z_h}}$.  Similar consequence
applies for the first group of solutions.
\begin{figure}[h]
\centering
\includegraphics[width=8 cm]{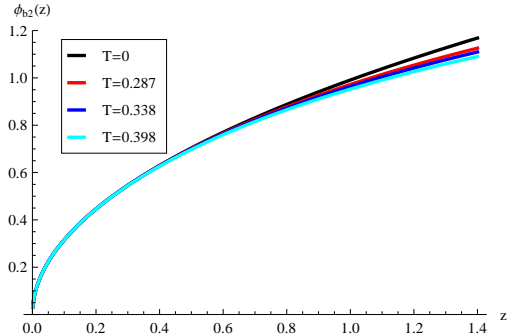}
\caption{The configuration of $\phi_{b2}(z)$ as a function
holographic coordinate $z$. Here we have fixed $p_{1\over 2}=1
\text{GeV}^{1\over 2}$ all over the paper, and the four curves'
parameters is as following {\{${T=0, f_{42}=0, p_{7\over 2}=0 }$\},
\{${T=0.287 \text{GeV}, f_{42}=0.5 \text{GeV}^4, p_{7\over
2}=-0.040...\text{GeV}^{7\over 2}}$\}, \{${T=0.338 \text{GeV},
f_{42}=1.0 \text{GeV}^4, p_{7\over 2}=-0.070... \text{GeV}^{7\over
2}}$\}, \{${T=0.398 \text{GeV} ,f_{42}=2.0 \text{GeV}^4, p_{7\over
2}=-0.121... \text{GeV}^{7\over 2}}$\}}. To get a smooth function
numerically, $p_{7\over 2}$ have very high-precision and we replace
the more additional number with "..." for short.}\label{phi-4}
\end{figure}

\begin{figure}[h]
\centering
\includegraphics[width=8 cm]{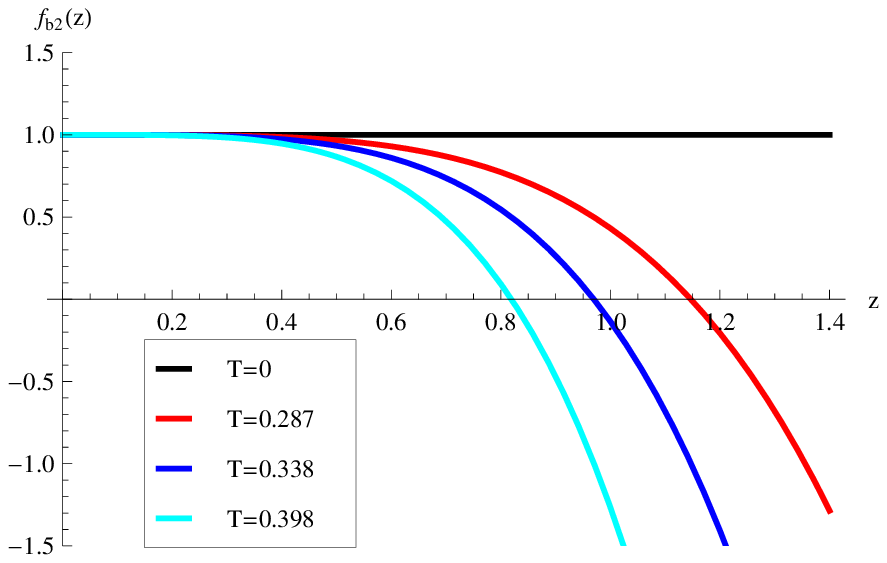}
\caption{The configuration of $f_{b2}(z)$ as a function holographic
coordinate $z$. Here we have fixed $p_{1\over 2}=1
\text{GeV}^{1\over 2}$ all over the paper, and the four curves'
parameters is as following \{${T=0, f_{42}=0, p_{7\over 2}=0 }$\},
\{${T=0.287 \text{GeV}, f_{42}=0.5 \text{GeV}^4, p_{7\over
2}=-0.040...\text{GeV}^{7\over 2}}$\}, \{${T=0.338 \text{GeV},
f_{42}=1.0 \text{GeV}^4, p_{7\over 2}=-0.070... \text{GeV}^{7\over
2}}$\}, \{${T=0.398 \text{GeV} ,f_{42}=2.0 \text{GeV}^4, p_{7\over
2}=-0.121... \text{GeV}^{7\over 2}}$\}. To get a smooth function
numerically, $p_{7\over 2}$ have very high-precision and we replace
the more additional number with "..." for short.}\label{ff-4}
\end{figure}

\begin{figure}[h]
\centering
\includegraphics[width=8 cm]{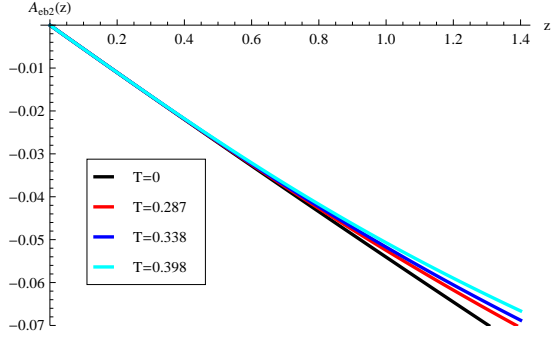}
\caption{The configuration of $A_{eb2}(z)$ as a function holographic
coordinate $z$.  Here we have fixed $p_{1\over 2}=1
\text{GeV}^{1\over 2}$ all over the paper, and the four curves'
parameters is as following \{${T=0, f_{42}=0, p_{7\over 2}=0 }$\},
\{${T=0.287 \text{GeV}, f_{42}=0.5 \text{GeV}^4, p_{7\over
2}=-0.040...\text{GeV}^{7\over 2}}$\}, \{${T=0.338 \text{GeV},
f_{42}=1.0 \text{GeV}^4, p_{7\over 2}=-0.070... \text{GeV}^{7\over
2}}$\}, \{${T=0.398 \text{GeV} ,f_{42}=2.0 \text{GeV}^4, p_{7\over
2}=-0.121... \text{GeV}^{7\over 2}}$\}. To get a smooth function
numerically, $p_{7\over 2}$ have very high-precision and we replace
the more additional number with "..." for short.} \label{Ae-4}
\end{figure}
\section{Energy momentum tensor and free energy}
In this section, we would like to study the stability of thermal gas
solutions and Euclidean black hole solutions by comparing the free
energy. Here we should stress that thermal gas solution is obtained
by Euclidean version of zero temperature solution as mentioned
previously \cite{Gursoy:2008za}. To obtain reasonable energy
momentum tensor on the boundary, one should introduce the suitable
counter terms. For later use, we just focus on these two groups of
thermal gas solutions and black hole solutions. One will see these
two groups of solutions capture different characters of UV
behaviors. These characters will lead to different behaviors of
entanglement temperature which will be studied in the next section.

\subsection{Energy momentum tensor}
In this subsection, we would like to introduce the counter terms to
cancel the divergent of the action and make the energy momentum
tensor of dual field theory well defined. In our cases, one can
find that we have to introduce $\phi^2, \phi^4,\phi^6$ terms to cancel
the divergences. For $\phi^2$ term, it is introduced as standard
AdS/CFT dictionary to make the energy momentum tensor of dual field
theory be well defined. These additional terms related to $\phi^4,
\phi^6$ will be helpful to give to well defined boundary energy
momentum tensor for the second black hole solution. The total action now becomes
\begin{eqnarray}\label{C2}
I_{\text{ren}}&=&S_{\text{5D}}+
S_{\text{GH}}+S_{\text{count}}\nonumber\\&=&\frac{1}{16 \pi G_5}
\int_M d^5
x\sqrt{-g^E}\left(R-\frac{4}{3}\partial_{\mu}\phi\partial^{\mu}\phi-V_E(\phi)
-\frac{Z(\phi)}{4g_{g}^2}F_{\mu\nu}F^{\mu\nu}\right)\nonumber \\&-&
\frac{1}{16\pi G_5} \int_{\partial
M}d^4x\sqrt{-\gamma}\Big[2K-{6\over L}+{8 \lambda_2 \phi^2 \over 3 L
}+{64 \lambda_4 \phi^4 \over 9 L^2 }
+{512 \lambda_6 \phi^6 \over 81 L^3 } \Big],\nonumber \\
\end{eqnarray}
with $\lambda_2, \lambda_4, \lambda_6$ are coefficients of count
terms $\phi^2, \phi^4, \phi^6 $ introduced here. These coefficients
can be fixed by canceling the divergences of boundary momentum
tensor. Here $K_{ij}$ and $K$ are respectively the extrinsic
curvature and its trace of the boundary $\partial M$, $\gamma_{ij}$
is the induced metric on the boundary $\partial M$. These quantities
are defined as follows
\begin{eqnarray}
\gamma_{\mu \nu}&=&g_{\mu \nu} +n_{\mu} n_{\nu},\\
K_{\mu\nu}&=&h^\lambda_\nu D_{\lambda} n_{\mu},\\
\gamma&=&\det(\gamma_{\mu\nu}),\\
K&=&g^{\mu\nu}K_{\mu\nu},
\end{eqnarray}
where $\gamma_{\mu \nu}$ denotes the induced metric, $n_{\mu}$
stands for the normal direction to the boundary surface $\partial M$
as well as $D_{\lambda}$ stands for covariant derivative.

In the asymptotical AdS space, the boundary surface locates at $z=0$
surface, and usually one has to regularized it to a finite
$z=\epsilon$ surface. So we have the normalized normal vector
$n_\mu=\frac{\delta^\mu_z}{\sqrt{g_{zz}}}$.

The first term of the last line in (\ref{C2}) is Gibbons-Hawking
term $S_{\text{GH}}$ and the remain terms are related to counter
terms $S_{\text{count}}$ related to cosmological constant and dilaton
field.

To regulate the theory, we restrict to the region $z\ge \epsilon$
and the surface term is evaluated at $z=\epsilon$. The induced
metric is $\gamma_{ij}=\frac{\tilde{L}^2}{\epsilon^2}
g_{ij}(x,\epsilon)$, where the leading term of expansion of
$g_{ij}(x,\epsilon)$ with respect to $\epsilon$ is the flat metric
$g_{(0)}^{ij}$. Then the one point function of stress-energy tensor
of the dual CFT is given by \cite{KS}\cite{SKS}
\begin{eqnarray}
T_{ij}=\frac{2}{\sqrt{-\det g_{(0)}}}\frac{\delta I_{ren}}{\delta
g_{(0)}^{ij}}=\lim_{\epsilon \to
0}\Big(\frac{{L}^2}{\epsilon^2}\frac{2}{\sqrt{-\gamma}}\frac{\delta
I_{ren}}{\delta \gamma^{ij}}\Big). \label{CFTET}
\end{eqnarray}
The finite part of boundary
energy-stress tensor is from the
$O(\epsilon^2)$ of the Brown-York tensor $T^{BY}_{ij}$ on the
boundary $z=\epsilon$, with
\begin{eqnarray}\label{BY}
T^{BY}_{ij}=-\frac{1}{16 \pi G_5}\Big[(K_{ij}-(K+{d-2\over
L}+{\lambda_2\over L}\phi(\epsilon)^2+{\lambda_4\over
L^2}\phi(\epsilon)^4+{\lambda_6\over
L^3}\phi(\epsilon)^6)\gamma_{ij})\Big],
\end{eqnarray}

In the first black solution, the coefficients of count terms can be
following\begin{eqnarray}
\lambda_2&=&{1\over 4},\nonumber\\
\lambda_4&=&0,\nonumber\\ \lambda_6&=&0.
\end{eqnarray}

Directly evaluate (\ref{BY}) using (\ref{CFTET}), we get \bea
T_{tt}={L^3\over 16 \pi G_5}{ }  ({3\over 2} f_{41} - p_1 ({2
p_1^3\over 81} + {2\over 3} p_3)). \label{ETGB}\eea One can see we
can only introduce the $\phi^2$ term to cancel the action to make
the boundary energy momentum tensor be well defined. These higher
powers of $\phi$, such as $\phi^4 , \phi^6$ terms, are not necessary
to be included.

The $tt$ component of energy tensor on the UV boundary of the second
black hole solution \bea T_{tt}={L^3\over 16 \pi G_5}
\left(\frac{3f_{42}}{2}-\frac{ p_{{1\over
2}}^8}{5511240}-\frac{p_{{1\over 2}} p_{{7\over 2}}}{2}\right).
\label{ETGB-second}\eea To obtain the well defined boundary energy
momentum tensor, it is necessary to introduce $\phi^2, \phi^4,
\phi^6$ terms. The coefficients of these terms can be determined by
canceling the divergences of boundary energy momentum tensor. In
this case, one should introduce the additional term $\phi^4, \phi^6$
terms to obtain the finite boundary momentum tensor. This aspect is
different from the previous case. Here we list the coefficients of
related counter terms
\begin{eqnarray}
\lambda_2&=&{1\over 8},\nonumber\\
\lambda_4&=&{L\over 1152},\nonumber\\ \lambda_6&=&{L^2\over 414720}.
\end{eqnarray} In terms of the results, one can introduce more
powers of $\phi$ terms as a strategy to cancel the divergences of
boundary energy momentum tensor for more general Einstein dilaton
system. Here we just only take two groups of solutions as examples
to show how to obtain the finite energy momentum tensor on the
boundary.

\subsection{The difference of free energy} In this subsection, we would like to study the
difference of free energy between thermal gas and Euclidean black
hole\footnote{In this subsection, all studies are based on Euclidean
version of gravity solution. In this paper, we denote thermal gas
solutions as Euclidean version of zero temperature solutions.}. In
terms of total Euclidean action given in (\ref{C2}), we can obtain
on-shell action for black hole as following
\bea\label{blackholeonshell} S_{\text{5D-BH}}&=&2\int d^3 x
\int^{\beta}_0 d\tau \int^{z_h}_0 dz
\partial_z(b_{eb}^2 b_{eb}^{'} f_{b})=2 V_3\beta (b_{eb}^2 b_{eb}^{'}
f)|_\epsilon^{z_h}\nonumber\\S_{\text{GH-BH}}&=&V_3 \beta\Big(
b_{eb}^{3}(\epsilon)f'_b(\epsilon) +8
b_{eb}^{2}(\epsilon)b'_{eb}(\epsilon)f_{b}(\epsilon)\Big) \eea with
$ b_{eb}(z)={e^{A_{eb}(z)}\over z}={e^{A_{sb}(z)-{2\over
3}\phi_{b}(z)}\over z}$, the period of Euclidean time ${\beta}=
{1\over T}$ and volume of space ${V_3}$. Here $S_{\text{BHcount}}$
is given by (\ref{C2}) with insertion of thermal gas solution and
$\epsilon$ is the regularized point. Here we also used
eqs.(\ref{onshellaction}) and (\ref{bbV}) to obtain
(\ref{blackholeonshell}). Due to that $\phi_b$ satisfies the
asymptotical AdS boundary condition, $S_{\text{BHcount}}=0$. Hence
it is not necessary to consider the contribution from
$S_{\text{BHcount}}$.

\begin{figure}[h]
\begin{center}
\epsfxsize=6.5 cm \epsfysize=5.5 cm \epsfbox{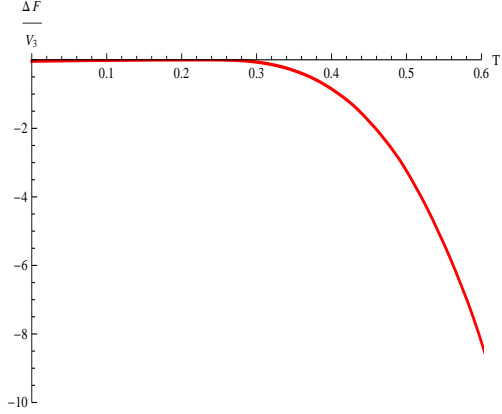}
\end{center}
\caption{The difference of free energy of black hole and thermal gas
in the first group solution. Here we choose the parameter $p_1=1.5
\text{GeV}$. }\label{freeenergy-3}
\end{figure}

\begin{figure}[h]
\begin{center}
\epsfxsize=6.5 cm \epsfysize=5.5 cm \epsfbox{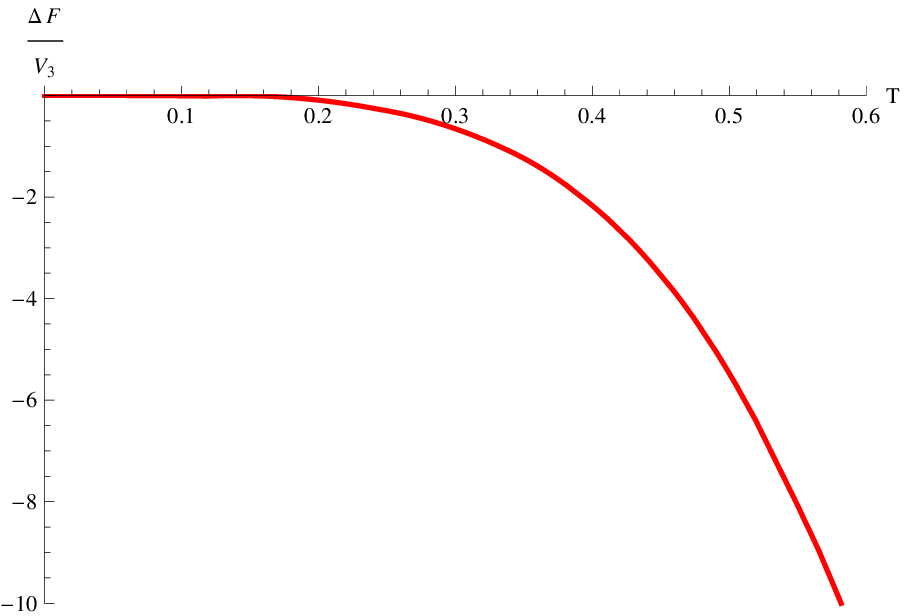}
\end{center}
\caption{The difference of free energy of black hole and thermal gas
in the second group. Here we choose the parameter $p_{1\over 2}=1
\text{GeV}^{1\over 2}$.}\label{freeenergy-4}
\end{figure}

In term of (\ref{blackholeonshell}), one replace $A_{eb},f_{b}$ with
$A_{et},f_{t}$ to calculate the free energy of thermal gas which is
only dependent on UV behavior of conformal factor $b_{et}(z)$ in the
following way \bea S_{\text{TG}}^{\text{reg}}&=&S_{\text{5D-TG}}+
S_{\text{GH}}+S_{\text{TGcount}},\nonumber\\
&=&\frac{1}{16\pi G_5}\lim_{\tilde{\epsilon} \mapsto
0}\tilde{\beta}(\tilde{\epsilon})\tilde{V}_3(\tilde{\epsilon})
\Big(6
b_{et}^2(\tilde{\epsilon})b_{et}'(\tilde{\epsilon})\Big)+S_{\text{TGcount}}
\label{TGregularaction},\eea with $b_{et}={e^{A_{et}}\over z}$, the
period of Euclidean  time $\tilde{\beta}= {1\over T}$ and volume of
space $\tilde{V_3}$. Where $S_{\text{TGcount}}$ is given by
(\ref{C2}) with insertion of thermal gas solution. We have checked
that for thermal gas solution, the integral of the Einstein-Hilbert
action extends on the region $(0, z_{IR})$, where $z_{IR}$ is the IR
cutoff and the IR contribution vanishes whenever $6
b_{et}^2(z_{IR})b_{et}'(z_{IR}) \rightarrow 0 $ as $ z\rightarrow
z_{IR} $. For these two thermal gas solutions, the IR contribution
vanishes as $ z\rightarrow z_{IR} $. Due to the fact that $\phi_{t}$
satisfy with the asymptotical AdS boundary condition, $
S_{\text{TGcount}}=0$. Here it is not necessary to consider the
contribution from $S_{\text{TGcount}}$.

In order to compare the free energy between black hole and thermal
gas, we should match the following conditions\cite{Gursoy:2008za}
\bea
b_{et}(\tilde{\epsilon})\tilde{\beta}(\tilde{\epsilon})=b_{eb}(\epsilon)\beta(\epsilon)\sqrt{g(\epsilon)},
\text{ }\text{  }\text{ }\text{ }{b}_{et}(\tilde{\epsilon})^3
\tilde{V}_3(\tilde{\epsilon})= b_{eb}(\epsilon)^3V_3(\epsilon).
\label{epsilon-epsilon}\eea Here $\epsilon,\tilde{\epsilon}$ are
different UV cutoff in black hole and thermal gas solution
respectively. One should match
$\phi_{th}(\tilde{\epsilon})=\phi_{b}(\epsilon)$ to obtain the
relationship between $\epsilon $ and $ \tilde{\epsilon}$. In terms
of the relationship, one can obtain the difference of on shell
action between thermal gas and black hole \bea
S_{\text{BH-TG}}&=&-\frac{\beta V_3}{16\pi
G_5}\Big(\lim_{\epsilon\mapsto0} \Big(
b_{eb}^{3}(\epsilon)f'_b(\epsilon)
+6b_{eb}^{2}(\epsilon)b'_{eb}(\epsilon)f_{b}(\epsilon)\Big)+S_{\text{BHcount}}\Big)\nonumber\\&&+\Big({b_{eb}^4(\epsilon)\sqrt{f_{b}(\epsilon)}\over
b_{et}^4(\tilde{\epsilon})}6
b_{et}^2(\tilde{\epsilon})b_{et}'(\tilde{\epsilon})+S_{\text{TGcount}}\Big).\label{onshellaction1}\eea
One can see the formula (\ref{onshellaction1}) \footnote{Here we use
the Euclidean action to calculate the difference of free energy. Our
results is consistent with \cite{Gursoy:2008za}.} is consistent with
on shell action \cite{Cai:2012xh} in EDM system with $U(1)$ gauge
field turned off. The difference of free energy $\beta \Delta F=
S_{\text{BH-TG}}$ between thermal gas and black hole are always
negative as shown in Fig.~\ref{freeenergy-3} and \ref{freeenergy-4}
which correspond to the first and second group of solutions
respectively. Here the $\Delta F$ denotes the difference of free
energies between them. In Fig.~\ref{freeenergy-3} and
\ref{freeenergy-4}, we just set parameters to show the difference of
free energy between black hole and thermal gas. As we
can see in Fig.~\ref{freeenergy-3} and \ref{freeenergy-4}, the black
hole solutions will go back to thermal gas solution when the
temperature decreases to zero. In the whole region $T\geq 0$, black
holes solutions are favored in these two groups of solutions. Here
one should note that these thermal gas solutions are obtained by
compactifying the time direction into a circle without considering
back reaction of equation of state of thermal gas. Therefore, black
hole phases is more stable than thermal gas phases in these two
groups of gravity solutions as shown in Fig.~\ref{freeenergy-3} and
\ref{freeenergy-4} in this level. We also
wound like to mention that in Fig.~\ref{freeenergy-3} and
\ref{freeenergy-4}, $\Delta F$ is always negative although its
absolute value is very small when $T$ is not large enough.

\section{Entanglement temperature}
As applications of these solutions from AdS/CFT point of view, we
 consider the novel quantity called entanglement temperature
in non-conformal cases
\cite{Klebanov:2007ws}\cite{Ryu:2006ef}\cite{Narayan:2013qga} for
our solutions. It is highly nontrivial to consider the dynamical
entanglement temperature. Now we have generated two zero temperature
solutions and the corresponding black hole solutions. Previous
studies in above sections have shown that the black hole solutions
can go back to zero temperature solutions. In this sense, that is to
say black hole solutions are thermal states of zero temperature
solutions in our cases. It is necessary to check whether there
exists the first law of thermodynamics for EE in this system.
\subsection{Variation of entanglement entropy in strip case}
In this subsection, we consider the subsystem with a stripe profile
which is defined by $-\frac{L_s}{2}<x_1\equiv x<\frac{L_s}{2}$ and
$-\frac{R_0}{2}<x_2,x_3<\frac{R_0}{2}$ where $R_0$, as a cutoff, is
the lengthes of $x_2, x_3$ direction.  Then the induced metric
$h_{\mu\nu}$ of the bulk surface after perturbation (or thermal
excitation) in the Einstein frame is
 \bea {L^2 e^{2A_e(z)}\over z^2
}\Big({1\over f(z)}+x'(z)^2\Big)dz^2+\Big( {L^2 e^{2A_e(z)}\over z^2
}\Big) dx_1^2+ \Big( {L^2 e^{2A_e(z)}\over z^2 }\Big)
dx_2^2,\nonumber\\\eea where $x$ is a function of $z$, $x=x(z)$ and
the prime stands for the derivative with respect to $z$ in this
subsection. We get the following volume of the submanifold which can
thought as an action \bea S= {L^3 R_0^2 \over 16 \pi G_5}\int dz
{1\over z^3} \sqrt{{e^{6 A_e(z)}(1+f(z)x'(z)^2)\over
f(z)}}.\label{EESF}\eea By minimizing the functional ($\ref{EESF}$)
to obtain the classical configuration, we get the following equation
of motion \bea
x''(z)&+&\frac{f (z)  x'(z)^3 \left(3 z A_{e}'(z)-3\right)}{z}\nonumber\\
&+&\frac{ x'(z)\left(2 f( z) \left(3z A_e'(z)-3\right)+z
f'(z)\right)}{2 z f(z)} =0.\eea

The solution is \bea \label{sublength} L_{s}(z_0)&=&2\int_0^{z_0}
{z^3\over \sqrt{(e^{6A_e(z)-6A_e(z_0)}z_0^6-z^6)f(z)}}
dz\nonumber\\&=& 2z_0 \int_0^{1} {t^3\over \sqrt{(e^{6A_e(z_0
t)-6A_e(z_0)}-t^6)f(z_0 t)}} dt,\eea where $z_0$ is the maximal
value of $z$ on the surface in the bulk, which is also called the
turning point. Here the turning point is defined by
$x'|_{z=z_0}=\infty$ and the boundary conditions give
 $x|_{z=0}=\pm L_s/2$.  We only care about  the case that the size $L_{s}$ of subsystem
 satisfies
$L_{s}\ll R_0$ which means $z_0 \ll R_0$. We have checked that
$L_s(z_0)$ is monotonic function of $z_0$ numerically. And one can
show that $z_0\ll R_0$ is equivalent to $L_s\ll R_0$.

Under this approximation, eq.~(\ref{sublength}) can be expressed by
following simple form. \bea L_s(z_0)=\frac{2\sqrt{\pi } \Gamma
\left(\frac{2}{3}\right)z_0}{\sqrt{f_{\text{b,th}}(0)} \Gamma
\left(\frac{1}{6}\right)}+ O(z_0),\eea where $f(z)=f_{\text{b,th}}$
correspond to configuration of black hole and zero temperature
solution respectively. We have expanded  the integrand in
(\ref{sublength}) in power series of $z_0$ and performed the
integration to obtain the $L_s(z_0)$ in case of $z_0 \ll R_0$.

The entanglement entropy in zero temperature solution is
\bea\label{eeth} S_{th}&=&\frac{L^3 R_0^2 }{8\pi G_5}\int_{0}^{z_0}
{dz\over z^3} \sqrt{\frac{z_0^6 e^{12 A_{et}(z)}}{f_{th}(z)
\left(z_0^6 e^{6 f_{t}(z)}-z^6 e^{6
A_{et}(z_0)}\right)}}\nonumber\\&=& \frac{ L^3 R_0^2 }{8\pi G_5
z_0^2}\int_{0}^{1} {dt\over t^3} \sqrt{\frac{z_0^6 e^{12 A_{et}(z_0
t)}}{f_{t}(z_0 t) \left(z_0^6 e^{6 f_{t}(z_0 t)}-(z_0 t)^6 e^{6
A_{et}(z_0)}\right)}}\eea

The entanglement entropy $S_{b}$ in black hole can be obtained by
replace $ A_{eth},  \phi_{th}, f_{th}$ with $ A_{eb},  \phi_{b},
f_{b}$. Here $S_{th}, S_{b}$  have not been regularized. If one is
only interested in the variation of entanglement entropy, one can
find that the two integrands in $S_{th}, S_{b}$   have the same
behavior near $z\sim 0$.

Now we introduce the counter terms for $S_{th}$ to resolve the
divergent of integrand at $t=0$, \bea
S_{\text{thEEcout}}&=&\frac{L^3 R_0^2 }{8\pi G_5}\int_0^1
dt\Big(\frac{e^{3 A_{et}(0)}}{\sqrt{f_{t}(0)} t^3 z_0^2}+\frac{e^{3
A_{et}(0)} \left(6 f_{t}(0) A_{et}'(0)-f_{t}'(0)\right)}{2
f_{t}(0)^{3/2} t^2 z_0}\nonumber\\&+&\frac{e^{3 A_{et}(0)} \left(-12
f_{t}(0) f_{t}'(0) A_{et}'(0)+36 f_{t}(0)^2 A_{et}'(0){}^2\right)}{8
f_{t}(0)^{5/2} t}\nonumber\\&+&\frac{e^{3 A_{et}(0)} \left(12
f_{t}(0)^2 A_{et}''(0)-2 f_{t}(0) f_{t}''(0)+3 f_{t}'(0)^2\right)}{8
f_{t}(0)^{5/2} t}\Big)\nonumber\\&+&\Big(-\frac{e^{3 A_{et}(0)}
\left(6 f_{t}(0) z_0 A_{et}'(0)-z_0 f_{t}'(0)+f_{t}(0)\right)}{2
f_{t}(0)^{3/2} z_0^2}\Big)\label{entanglementcount}\eea Where the
final term comes from compensation to deal with divergent. Combining
(\ref{eeth}) and (\ref{entanglementcount}) will lead to
$S^{reg}_{th}=S_{th}-S_{\text{thEEcout}}$ with  putting the
geometrical functions $A_{et}=A_{etn}, f_{t}=f_{tn},
\phi_{t}=\phi_{tn}$ with $n=1,2$.\footnote{In this paper, we just
use $A_{ebn}, f_{bn}, \phi_{bn}$ to denote the $n$-th black hole
solutions and $A_{etn}, f_{tn}, \phi_{tn}$ stand for $n$-th zero
temperature solutions. } In term of (\ref{eeth}) and
(\ref{entanglementcount}), one just only replace $A_{et},f_{t}$ with
$A_{eb},f_{b}$ to obtain $S^{reg}_{b}$. Finally, the variation of
entanglement entropy \bea \label{difference-BHTG}\Delta S=
S^{reg}_{b}-S^{reg}_{th}.\eea In the following part, we consider the
black hole as low thermal excitation of zero temperature solution
and calculate the difference of entanglement entropy between them
approximately.

\subsection{Entanglement temperature in strip case}
\cite{Bhattacharya:2012mi} has proposed a universal relation between
the variance of the energy and the entanglement entropy for a small
subsystem on the boundary theory. The universal relation induce a
novel concept called entanglement temperature. The $t-t$ component
of the energy-stress tensor \cite{KS}\cite{SKS} corresponds to the
energy density (\ref{ETGB}) and (\ref{ETGB-second}) in first and
second black hole solution respectively.

For the finite stripe, the variance of entanglement entropy
(\ref{difference-BHTG}) in the subsystem of the first group of
gravity solutions is \bea \label{entanglementfst}\Delta S_{fst}&=&
{(0.350546 f_{41} - 0.409903 p_1 p_3) L^3 R_0^2\over 8 \pi G_5 }
z_0\nonumber\\&=& {(0.350546 f_{41} - 0.409903 p_1 p_3) L^3
R_0^2\over 8 \pi G_5 }{L_s^{{2}} \Gamma ({1\over 6})^{{2}} \over
\pi^{{2}} \Gamma ({2\over 3})^{{2}}}.\eea Where $p_{t1}=p_{b1}=p_1,
f_{t41}=0, p_3=0 $ and $f_{b41}=f_{41}, p_{b3}= p_3$ \footnote{To
make the notation clear, $p_{t1}, f_{t41}, p_{t3} $ and $p_{b1},
f_{b41}, p_{b3} $ denote the coefficients of thermal gas (or the
zero temperature solution) and black hole respectively in the first
group of solutions.}. On the other hand, the increased amount of
energy in the subsystem with strip configuration is given by \bea
\Delta E&=& \int
dx^3\left(T_{tt}^{b1}-T_{tt}^{t1}\right),\nonumber\\&=&{L^3 R_0^2
L_s\over 8 \pi G_5}{ } ({3\over 2} f_{41} -   {2\over 3}
p_1p_3).\eea Where $T_{tt}^{b1}$ and $T_{tt}^{t1}$ stand for the
$tt$ component of boundary energy momentum tensor in black hole and
zero temperature in first group solutions respectively. For the
exact formula about $ T_{tt}$, we refer to eq.~(\ref{ETGB}). For
$p_1=0$, the first zero temperature solution and the first black
hole solution will go back to the case \cite{Bhattacharya:2012mi}
and the variation of entanglement entropy \footnote{The variation of
entanglement entropy (\ref{entanglementfst})(\ref{entanglementsnd})
is consistent with \cite{Bhattacharya:2012mi} up to normalization
factor in $p_1=p_{1\over 2}=0$ cases.} and boundary energy momentum
tensor also reproduce boundary momentum tensor given in
\cite{Bhattacharya:2012mi}.

The entanglement temperature in the first black hole is
\begin{eqnarray}\label{universal-relation-GB-stripe}
\frac{1}{T_{ent}}&=&{\Delta S_{fst} \over \Delta
E_{fst}}\nonumber\\&=&{(0.350546 f_{41}  - 0.409903 p_1 p_3) \over
({3\over 2} f_{41} - {2\over 3}p_1  p_3)}  {L_s \Gamma ({1\over
6})^{{2}} \over \pi^{{2}} \Gamma ({2\over 3})^{{2}}}.
\end{eqnarray} One
can find that the ${T_{ent}} \sim [E]$ through dimensional analysis
with $[f_{41}]= [E^{4}],[p_1]=[E^{1}],[p_3]=[E^{3}], [L_{s}]=
[E^{-1}]$. Here $R_0$ is considered as the character length of the
subsystem on the boundary. In terms of the dimensional analysis, one
can see the coefficient highly depends on the geometry and shape of
strip. In the first group of solutions, we fix the $p_1=1.5 GeV$ for
convenience to study the entanglement temperature without loss of
generality. From our numerical study, $f_{41}, p_3$ are not
independent and they are fixed by horizon condition $f(z_h)=0$. In
this sense, $f_{41}, p_3$ are functions of the position of black
hole horizon $z_h$ or black hole temperature. In our cases, we just
only turn on small $f_{41}$ or $p_{3}$ to control the temperature
shown in Fig.~\ref{f4-T-3}. 
The temperature $T={{f_{b1}(z)'\over 4\pi}|_{z=z_h}}$ increases with
$f_{41}$ and  $p_3$ decreases simultaneously in the first black hole
solution. The temperature goes from zero to small positive value
which corresponds to that $f_{41}$ increases from zero to small
positive value. At the same time, the behavior of $p_3$ decreases
from zero to negative value monotonically. In terms of
(\ref{universal-relation-GB-stripe}), both of the nominator and
denominator in (\ref{universal-relation-GB-stripe}) are all positive
\footnote{We have checked this statement for $p_1\geq 0$
numerically. It is hard to prove it analytically due to the relation
between $f_{41}$ and $p_3$ controlled by black hole horizon at this
stage.} and the entanglement temperature are positive which is
consistent with thermodynamical first law like proposed by
\cite{Bhattacharya:2012mi}.

\begin{figure}[h]
\begin{center}
\epsfxsize=6.5 cm \epsfysize=5.5 cm \epsfbox{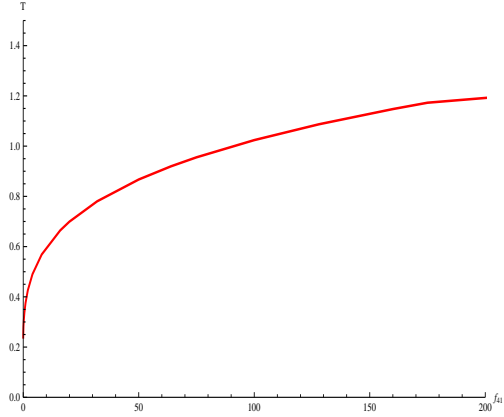}
\end{center}
\caption{T v.s. $f_{41}$  in the first solution when $p_1=1.5
\text{GeV}$. }\label{f4-T-3}
\end{figure}

For the finite stripe, the variance of entanglement entropy
(\ref{difference-BHTG}) in the subsystem of the second background is
\bea \label{entanglementsnd}\Delta S_{snd}&=&{(0.350546 f_{42} -
0.23911 p_{{7\over 2}} p_{1\over 2} ) L^3 R_0^2\over 8 \pi G_5 }
z_0^{{3\over 2}}\nonumber\\&=& {(0.350546 f_{42} - 0.23911
p_{{7\over 2}} p_{1\over 2} ) L^3 R_0^2\over 8 \pi G_5 }{L_s^{{2}}
\Gamma ({1\over 6})^{{2}} \over \pi^{{2}} \Gamma ({2\over
3})^{{2}}}.\eea Where $p_{t2}=p_{b21}=p_2, f_{t42}=0, p_{7t\over
2}=0
 $ and $f_{b42}=f_{42}, p_{7b\over 2}= p_{7\over
2}$ \footnote{ $p_{1t\over 2}, f_{t42}, p_{7t\over 2} $ and
$p_{1b\over 2}, f_{b42}, p_{7b\over 2} $ denote the coefficients of
the zero temperature solution and the black hole respectively in the
second group of solutions.}. The increased amount of energy in the
subsystem with strip configuration is given by \bea \Delta E&=& \int
dx^3\left( T_{tt}^{b2}-T_{tt}^{t2}\right),\nonumber\\&=&{L^3 R_0^2
L_s\over 8 \pi G_5} \left(\frac{3f_{42}}{2}-\frac{p_{{1\over 2}}
p_{{7\over 2}}}{2}\right).\eea Where $T_{tt}^{b2}$ and $T_{tt}^{t2}$
stand for the $tt$ component of boundary energy momentum tensor in
black hole and zero temperature solution in second group solutions
respectively. The exact formula about $ T_{tt}$ is given in
eq.~(\ref{ETGB-second}). For $p_{1\over 2}=0$, the second zero
temperature solution will go back to case \cite{Bhattacharya:2012mi}
and the variation of entanglement entropy and boundary energy
momentum tensor are also consistent with boundary momentum tensor
given in \cite{Bhattacharya:2012mi}.

The entanglement temperature in the second black hole is
\begin{eqnarray}\label{universal-relation-GB-stripe-Second}
\frac{1}{T_{ent}}&=&{\Delta S_{snd}\over \Delta
E_{snd}}\nonumber\\&=&{(0.350546 f_{42} - 0.23911 p_{{7\over 2}}
p_{1\over 2} ) \over  \left( \frac{3f_{42}}{2}-\frac{p_{{1\over 2}}
p_{{7\over 2}}}{2}\right)}{L_s \Gamma ({1\over 6})^{{2}} \over
\pi^{{2}} \Gamma ({2\over 3})^{{2}}}
\end{eqnarray}
One can also find that the ${T_{ent}} \sim [E]$ through dimensional
analysis with $[f_{42}] =[E^{4}],[\mu] =[E^{1}],[p_n]= [E^{n}],
[L_{s}]=[E^{-1}]$. In the second group of solutions, we fix the
$p_{1\over 2}=1\text{GeV}$ for convenience to study the entanglement
temperature without losing generality. From our numerical study,
$f_{42}, p_{7\over 2}$ are not independent and they are fixed by
horizon condition $f_{b1}(z_h)=0$. In this sense, $f_{42}, p_{7\over
2}$ are functions of the position of black hole horizon $z_h$ or
black hole temperature. In our cases, we turn on small $ f_{42}$ or
$ p_{7\over 2}$ to control the temperature shown in figures
[\ref{f4-T-4}]. The black hole temperature $T={{f_{b2}(z)'\over
4\pi}|_{z=z_h}}$ increases with $f_{42}$ and $p_{7\over 2}$
decreases simultaneously in the second black hole solution. The
temperature goes from zero to small positive value which corresponds
to that $f_{42}$ increases from zero to small positive value. At the
same time, the behavior of $p_{7\over 2}$ decreases from zero to
negative value monotonically. In terms of
(\ref{universal-relation-GB-stripe-Second}), the nominator and
denominator of (\ref{universal-relation-GB-stripe-Second}) are all
positive \footnote{We also have checked this statement for
$p_{1\over 2} \geq 0$ numerically.} and the entanglement temperature
are positive which is the same as previous case.

\begin{figure}[h]
\begin{center}
\epsfxsize=6.5 cm \epsfysize=5.5 cm \epsfbox{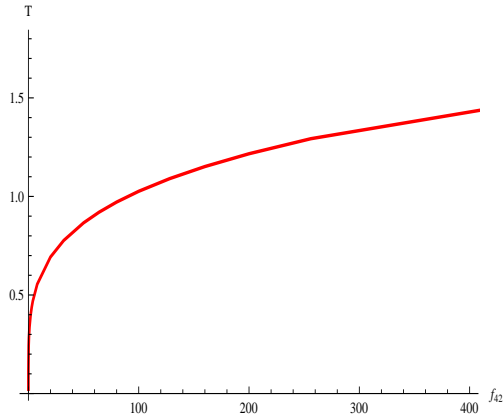}
\end{center}
\caption{T v.s. $f_{42}$ in the second solution when $p_{1\over 2}=1
\text{GeV}^{1\over 2}$.}\label{f4-T-4}
\end{figure}

One can see that the entanglement temperatures
(\ref{universal-relation-GB-stripe})
(\ref{universal-relation-GB-stripe-Second}) denote the first law
relation of thermodynamics for EE  in these two non-conformal cases
from holographic perspective. One can note that entanglement
temperature
(\ref{universal-relation-GB-stripe})(\ref{universal-relation-GB-stripe-Second})
can reproduce that given in \cite{Bhattacharya:2012mi} with
$p_1=p_{1\over 2}=0$. For non-conformal cases $p_1=p_{1\over 2}\neq
0$, entanglement temperatures are also related to both UV and IR
geometries which characterized by $p_1 \text{} \text{or} \text{}
p_{1\over 2}$ and $f_{41}, p_{3} \text{} \text{or} \text{} f_{42},
p_{7\over 2} $ respectively. In the first group of
solutions, the parameter $p_3$ is related to condensation of the
dimension $3$ operator $\mathcal {O}_{3}$ \footnote{In terms of
holographic dictionary $\Delta_{\pm}={1\over 2}(d\pm \sqrt{d^2+4
M^2})$, $M_{\phi_{b1}}^2=-3$ and $\langle\mathcal {O}_{3}\rangle
=\sqrt{{8\over 3}} p_3$. Where $d$ is dimension of field theory and
$M, M_{\phi_{b1}}$ are the normalized bulk mass of scalar field. At
same time, $M_{\phi_{b2}}^2=-{7\over 4}$ and $\langle\mathcal
{O}_{{7\over 2}}\rangle =\sqrt{{8\over 3}}
\frac{1}{653184}\left(653184 p_{\frac{7}{2}}-5
p_{\frac{1}{2}}^7\right) $ .} which holographically dual to scalar
$\phi_{b1}$ at special temperature. The exact relation can be read
out from asymptotic expansion of holographic coordinate $z$ near UV
region from (\ref{firstblackhole}) in terms of holography
dictionary. In this case, the temperature is determined by the
$f_{41}$ with fixing non-vanishing source $p_1$. In the second group
of solutions, due to the bulk mass of $\phi_{b2}$, $\frac{1
}{653184}\left(653184 p_{\frac{7}{2}}-5 p_{\frac{1}{2}}^7\right) $
corresponds to condensation of operator $\mathcal {O}_{7\over 2}$
with dimension ${7\over 2}$ living on the boundary. Where the
condensation is induced by the source $p_{1\over 2}$. From
(\ref{BHsolution2}), one can easily read out the relation between
the source and condensation of corresponding operator in the same
way. Here we still have no idea about exact physical meaning of
these operators partly due to that we take a bottom-up approach.  One should note that there should be two ways
quantize $\phi_{b1, b2}$ by imposing Dirichlet or Neumann conditions
at the aAdS boundary, which are often called standard and
alternative quantization respectively, and lead to two different
QFTs. The analogy analysis can be done in the same way and we do not
repeat here. Therefore, non-conformal entanglement temperatures do
not only depend on geometric data of the subsystem but also data of
gauge theory living on the boundary.

\section{Conclusion and discussion}
In this paper, motivated by studying the dynamics of entanglement
entropy from Einstein equation in ED system, we make use of novel
technology called potential reconstruction approach to generate
general gravity solutions in EDM system and we further study the
entanglement temperature with Maxwell field turned off. The
potential reconstruction provides an easy way to investigate novel
quantities such as entanglement temperature in complicated
non-conformal system from holographical point of view.

Firstly, we generate various gravity solutions within this approach
and one can find that the scalar potential appeared in total action
depends not only on configurations of fields but also on integral
constants, such as $f_{41}, f_{42}$ which are related to
temperature. This is an effective method to obtain gravity
solutions. For simplify our analysis, we fix the scalar potential to
be independent of any other parameters except cosmological constant.
Through some guess works, one can obtain some zero temperature
solutions as we show in section 3. In order to study entanglement
temperature, one should turn on thermal excitation of zero
temperature solution which corresponds to the black hole solution in
this paper. There is strong constrains that the black hole solution
can go back to zero temperature solution continuously by tuning some
temperature parameters $f_{bn}$. Therefore, to find black hole
solution in fixed scalar potential is not an easy job here. Here we
proposed a numerical way to find corresponding black hole solution
to avoid the parameters dependence of dilaton potential.

Secondly, in order to study the stability of the thermal gas
solution and Euclidean black hole solution, we also compute the free
energy of them through introducing finite powers of $\phi$ terms as
counter terms. In these two groups of solutions, the difference of
free energy between thermal gas and black hole shows that thermal
gas solution is more unstable than the corresponding black hole
solution. In this sense, we consider the black hole solution as an
stable thermal excitation of zero temperature solution. The
entanglement entropy is a candidate for entropy in non-equilibrium
physics. It is important to study the fundamental properties of
entanglement entropy in order to understand non-equilibrium physics.
In this paper, we have tuned some numerical parameter which is dual
to black hole temperature to excite zero temperature solution. After
the thermal excitation, we also study the holographic entanglement
temperature and check that the first law of thermodynamics for HEE
also  exists in these two groups of solutions dynamically.  This
study lead us to understanding of non-equilibrium physics as well.

In the future, we would like to study the entanglement temperature
and entanglement density in EDM system. Furthermore, it is also
worth to try and check whether the first and second laws for HEE are
correct in theories dual to  gravity  coupled with more general
matter fields and/or with quantum corrections included
\cite{Faulkner:2013ana}. Finally, the authors of
\cite{Li:2011hp}\cite{He:2011hw}\cite{Cai:2012xh}\cite{Cai:2012eh}
have used potential construction approach to generate some gravity
solutions analytically. We can also make use of our numerical
methods to obtain other phases. It is possible to study the
properties of these phases. In appendix B, we list various
asymptotical $AdS_5$ solutions and some of them are good places to
study gauge/gravity correspondence in the bottom-up approach.
\section*{Acknowledgements}
The authors are grateful to Rong-Gen Cai, Zhoujian Cao, Wu-Zhong
Guo, Mei Huang, Li Li, Hong Lu, Bin Qin,  Jun Tao, Yu Tian,
Jian-Feng Wu, Jie Yang, Jia-Ju Zhang for useful conversations and
correspondence. Further we should thank Tadashi Takayanagi for his
nice suggestions and comments on this version. This work was
supported in part by the National Natural Science Foundation of
China a (No.10821504 (SH), No.10975168 (SH), No.11035008 (SH),
No.11305235(SH), No. 11105154 (JW), and No. 11222549
(JW)), and in part by Shanghai Key Laboratory of Particle Physics
and Cosmology under grant No.11DZ2230700. SH also would like
appreciate the general financial support from China Postdoctoral
Science Foundation No. 2012M510562. JW gratefully acknowledges the
support of K. C. Wong Education Foundation and Youth Innovation
Promotion Association, CAS as well.

\appendix
\renewcommand{\theequation}{\thesection.\arabic{equation}}
\addcontentsline{toc}{section}{Appendices}
\section{Appendix: Search of the Black Hole Solution Numerically}
With  $U(1)$ gauge field turned off in
Eq.(\ref{minimal-Einstein-action}), the system will be reduced to
Einstein-dilaton system as following form
\begin{eqnarray}\label{SGD}
S_{GD}=\frac{1}{16 \pi G_5} \int d^5 x
\sqrt{-g^E}\left(R-\frac{4}{3}\partial_{\mu}\phi\partial^{\mu}\phi-V_E(\phi)\right)
\end{eqnarray}

The Einstein equation and the field equation for the dilaton field
are,

\begin{eqnarray}
E_{MN}+\frac{1}{2}g^E_{MN}\left(\frac{4}{3}\partial_l\phi\partial^l\phi+V_E(\phi)\right)-\frac{4}{3}\partial_M\phi\partial_N\phi=0,
\label{einstein-dilaton-eq}\\
\frac{8}{3\sqrt{g_E}}\partial_M(\sqrt{g_E}\partial^M\phi)-\partial_{\phi}V_E(\phi)=0, \label{dilaton-field-Eq}
\end{eqnarray}
 with $E_{MN}=R_{MN}-\frac{1}{2}g^E_{MN}R$.

Contracting all Lorentz indices in (\ref{einstein-dilaton-eq}), we
can obtain

\begin{eqnarray}
-\frac{3}{2}R+\frac{5}{2}(\frac{4}{3}\partial_l\phi\partial^l\phi+V_E(\phi))-\frac{4}{3}\partial_l\phi\partial^l\phi=0
\end{eqnarray}

Inserting this equation to Eq.(\ref{SGD}), we get the on-shell
action as
\begin{eqnarray}
S_{on-shell}=\int d^5x \sqrt{-g^E}\left(\frac{2}{3}V_E(\phi)\right).
\label{onshellaction}
\end{eqnarray}

Then we would like to find the black hole solution in such ED system
numerically. With taking the 4D symmetry and the asymptotically AdS
condition into account, the metric ansatz would be taken as

\begin{eqnarray}
dS^2=b_e^2(z)(-f(z)dt^2+dx_i dx^i+\frac{1}{f(z)}dz^2)
\end{eqnarray}

With this ansatz, the geometric quantities can be calculated as

\begin{eqnarray}
R_{tt}&=&\frac{2b_e^{'2}}{b_e^2}f^2+\frac{5b_e^{'}}{2b_e}f f^{'}+\frac{b_e^{''}}{b_e}f^2+\frac{1}{2}f f^{''}\\
R_{zz}&=&-\frac{4 b_e^{''}}{b_e}+\frac{4 b_e^{'2}}{b_e^{2}}-\frac{5 b_e^{'} }{2 b_e }\frac{f^{'}}{f}-\frac{f^{''}}{2 f(z)}\\
R_{ii}&=&-\frac{2 b_e^{'2}}{b_e^{2}}f -\frac{b_e^{'}}{b_e}f^{'}-\frac{b_e^{''}}{b_e}f=-\frac{\partial_z(b_e^{'}b_e^{2}f)}{b_e^3}\\
R&=&-\frac{8 f(z) b_e^{''}}{b_e^3}-\frac{8 b_e^{'}
f'(z)}{b_e^{3}}-\frac{4 f(z) b_e^{'2}}{b_e^{4}}-\frac{f^{''}}{b_e^2}
\end{eqnarray}

We assume that all the quantities depend on the holographic
direction $z$ only. Then the nontrivial component of
Eq.(\ref{einstein-dilaton-eq}) are only $tt$ $zz$ $ii$ components.
Together with Eq.(\ref{dilaton-field-Eq}), there're $4$ non-trivial
equations
\begin{eqnarray}
E_{tt}+\frac{1}{2}g_{tt}(\frac{4}{3}\phi^{'2}g^{zz}+V_E(\phi))&=&0\\
E_{zz}+\frac{1}{2}g_{zz}(\frac{4}{3}\phi^{'2}g^{zz}+V_E(\phi))-\frac{4}{3}\phi^{'2}&=&0\\
E_{ii}+\frac{1}{2}g_{ii}(\frac{4}{3}\phi^{'2}g^{zz}+V_E(\phi))&=&0\\
\frac{8}{3\sqrt{g}}\partial_z(\sqrt{g}g^{zz}\partial_z\phi)-\partial_{\phi}V_E(\phi)&=&0
\end{eqnarray}

These equations can be written as,

\begin{eqnarray}
\frac{2E_{tt}}{g_{tt}}+\frac{4}{3}\phi^{'2}g^{zz}+V_E(\phi)&=&0\\
\frac{2E_{ii}}{g_{ii}}+\frac{4}{3}\phi^{'2}g^{zz}+V_E(\phi)&=&0\\
\frac{2E_{zz}}{g_{zz}}-\frac{4}{3}\phi^{'2}g^{zz}+V_E(\phi)&=&0\\
\partial_\phi V_E(\phi)=\partial_zV_E(\phi)/\phi^{'}&=&\partial_z\left(-\frac{2E_{tt}}{g_{tt}}-\frac{4}{3}\phi^{'2}g^{zz}\right)/\phi^{'}
\end{eqnarray}

Then, we can further rearrange these equations as follows
\begin{eqnarray}
V_E(\phi)&=&\frac{E_{zz}}{g_{zz}}+\frac{E_{tt}}{g_{tt}}=-3 g^{ii}R_{ii}=-3\frac{\partial_z(b_e^{'}b_e^{2}f)}{b_e^5}\\
\frac{4}{3}\phi^{'2}g^{zz}&=&\frac{E_{zz}}{g_{zz}}-\frac{E_{tt}}{g_{tt}}=\frac{R_{zz}}{g_{zz}}-\frac{R_{tt}}{g_{tt}}=-3\frac{f}{b_e^2}(\frac{b_e^{''}}{b_e}-2\frac{b_e^{'2}}{b_e^2})\\
0&=&\frac{E_{ii}}{g_{ii}}-\frac{E_{tt}}{g_{tt}}=\frac{R_{ii}}{g_{ii}}-\frac{R_{tt}}{g_{tt}}=-\frac{1}{2b_e^2}(f^{''}+3\frac{b_e^{'}}{b_e}f^{'})
\end{eqnarray}

So the simplified equations of motion are,
\begin{eqnarray}
-\frac{b_e^{''}}{b_e}+2\frac{b_e^{'2}}{b_e^2}&=&\frac{4}{9}\phi^{'2}\label{bbphi}\\
f^{''}+3\frac{b_e^{'}}{b_e}f^{'}&=&0\label{bbf}\\
-3\frac{\partial_z(b_e^{'}b_e^{2}f)}{b_e^5}&=&V_E(\phi)\label{bbV}\\
\frac{8}{3b_e^5}\partial_z(b_e^3
f\partial_z\phi)-\partial_{\phi}V_E(\phi)&=&0\label{phifield}
\end{eqnarray}
One should note that these four equations are not independent.
(\ref{bbphi})(\ref{bbf}) are 2nd order differential equations. One
of (\ref{bbV}) and (\ref{phifield}) is constrain equation. In order
to set our numerical strategy, we choose
(\ref{bbphi})(\ref{bbf})(\ref{phifield}) to find numerical solution
and use (\ref{bbV}) as consistent condition to check the numerical
solution.  This can be understood from that
\begin{eqnarray}
\partial_z V_E(\phi)&&=\phi^{'}\partial_\phi V_E(\phi)=\frac{8\phi^{'}}{3\sqrt{g}}\partial_z(\sqrt{g}g^{zz}\phi^{'})\\
&&=\frac{1}{\sqrt{g}}\left(2\partial_z(\sqrt{g}g^{zz})(\frac{4}{3}\phi^{'2})+\sqrt{g}g^{zz}\partial_z(\frac{4}{3}\phi^{'2})\right)\\
&&=-\frac{3}{b_e^5}\left(2\partial_z(b_e^3f)(\frac{b_e^{''}}{b_e}-2\frac{b_e^{'2}}{b_e^2})+b_e^3f\partial_z(\frac{b_e^{''}}{b_e}-2\frac{b_e^{'2}}{b_e^2})\right)\\
&&=-3\partial_z(\frac{\partial_z(b_e^2b_e^{'}f)}{b_e^5})+\frac{3b_e^{'}(3b_e^{'}f^{'}+b_ef^{''})}{b_e^4}.
\end{eqnarray}
So from Eq.(\ref{bbphi})(\ref{bbf})(\ref{phifield}), we can get
\begin{equation}
\partial_z(3\frac{\partial_z(b_e^{'}b_e^2f)}{b_e^5}+V_E(\phi))=0
\end{equation}
and if the initial condition guarantees that Eq.(\ref{bbV}) is
satisfied at a certain $z$,   Eq.~(\ref{bbV}) would be satisfied for
all $z$. So the four equations are not independent. We would only
use Eq.(\ref{bbphi})(\ref{bbf})(\ref{phifield}) in the numeric
process, and guarantee that Eq.(\ref{bbV}) using the initial
condition. Finally, there are total five integral constants which
should be fixed. These five constants will be fixed later.

\subsection{The first numerical black hole solution}

We take $V_{E1}(\phi)=-\frac{12}{L^2}-\frac{9 \sinh ^2\left(\frac{2
\phi}{3}\right)}{L^2}$, the first analytic solution can be generated
by potential reconstruction approach
\begin{eqnarray}
\phi_{t1}(z)&=& p_1 z\\
A_{et1}(z)&=&\log\left(\frac{{2\over 3}p_1 z}{\sinh({2\over 3}p_1 z)}\right)\\
f_{t1}(z)&=&1.
\end{eqnarray}

Then we try to find a asymptotic AdS black hole solution
numerically. In order to show this algorithm to obtain numerical
solution, we assume the expansion of $\phi(z)$ as
\begin{equation}
\phi_{b1}(z)=p_\Delta z^{\Delta}+p_{4-\Delta} z^{4-\Delta}+...
\end{equation}
The power order of $z$ can be determine $\Delta=1$(or equivalently
$\Delta=3$) from  the mass term in $V(\phi)$. We use series
expansion of unknown functions as follows
\begin{eqnarray}
\phi_{b1}(z)&=&p_1 z+p_3 z^3+\Sigma_n p_n z^n\label{phiI}\\
A_{eb1}(z)&=&\Sigma_n a_n z^n\label{AEI}\\
f_{b1}(z)&=&1+\Sigma_n f_n z^n\label{fI}
\end{eqnarray}
One should note these power orders of $z$ for each unknown function
should be consistent with Einstein equations. The other parameters
$p_n, a_n, f_n$ \footnote{One should note that $f_4=f_{41}$ in this
case.} can be determined by equations of motion. Here $f_{b1}(0)$ is
set to be one to satisfy  the asymptotical AdS boundary.

In terms of the equation of motion, the series expansion can be
determined in terms of  coefficients $p_3,f_{41}$ which can be
considered as the
integral constants of these differential equations. 
In some sense, these coefficients $p_1,p_3,f_{41}$ stand for IR
boundary conditions. The results are as following, with the even
powers of $\phi_{b1}(z)$ and odd powers of $A_{eb1}(z),f_{b1}(z)$
being always vanishing,
\begin{eqnarray}
\phi_{b1}(z)&=&p_1 z+p_3 z^3+\frac{z^5 \left(405 f_{41} p_1+612 p_1^2 p_3\right)}{3240}+...\\
A_{eb1}(z)&=&-\frac{2 p_1^2 z^2}{27}+z^4 \left(\frac{4 p_1^4}{3645}-\frac{2 p_1 p_3}{15}\right)+...\\
f_{b1}(z)&=&1-f_{41} z^4-\frac{4}{27} f_{41} p_1^2 z^6+...
\end{eqnarray}

Without loss of generality, we would fix $p_1=\frac{3}{2}
\text{GeV}$ as a setting of the energy scale in all the
calculations. In order to get a black hole solution, we would try to
get a solution with a pole in $f_{b1}(z)$.  $f_{b1}(z)$ should
decrease monotonously from the initial value $f_{b1}(z=0)=1$ to
$f_{b1}(z_h)=0$. Where $z_h$ denotes the event horizon of the black
hole. Re-writing eq.(\ref{phifield}) in terms of $A_{eb1}$, it
becomes
\begin{eqnarray}
\phi_{b1}^{''}-(\frac{3}{z}-3A_{e1}^{'})\phi^{'}-\frac{-8f_{b1}^{'}z^2\phi_{b1}^{'}+3e^{2A_{eb1}}L^2\partial_{\phi_{b1}}
V_{Et1}(\phi)}{8z^2f_{b1}}=0,
\end{eqnarray}
and due to $f_{b1}(z_h)=0$, $z_h$ would be a singular point of the
equation. To resolve this issue, the solution should satisfy that
$-8f_{b1}^{'}z^2\phi_{b1}^{'}+3e^{2A_{eb1}}L^2\partial_{\phi_{b1}}
V_{Et1}(\phi)=0$ at $z=z_h$ and this condition would impose a
constrain on the acceptable value of the two integral constants
$p_3,f_{41}$ with  $p_1$ fixed, i.e. if we take a certain $f_{41}$,
only a certain value of $p_3=p_3(f_{41})$ can create a black hole
solution. Varying $f_{41}$ would be related to varying the
temperature of the black hole.

To show how the above procedure works explicitly, we take
$f_{41}=0.75$ as an example. For further convenience, we define a
$\text{Test}(z)$ function \bea \label{test}\text{Test}(z)\equiv
-8f_{b1}^{'}z^2\phi_{b1}^{'}+3e^{2A_{et1}}L^2\partial_{\phi_{b1}}
V_{Et1}(\phi).\eea

The shooting method can find the exact $z_h$, such that
$f_{b1}(z_h)=\text{Test}(z_h)=0$. Here we will fix $p_1=3/2,
f_{41}=0.75$ to show how to find $z_h$ in Fig.~\ref{tune-p3}(a)(b).

We choose  $p_3=-0.1$, and insert this three integral constants into
Eqs.~(\ref{phiI})(\ref{AEI})(\ref{fI}). Then we fix the $p_1={3\over
2}$ to find the exact relation between $p_3$ and $f_{41}$. The
relation can be fixed by IR boundary condition $f_b(z_h)=0$ and
$\text{Test}(z_h)=0$ simultaneously. Here we can use shoot method to
find the exact $z_h$ and then $p_3, f_{41}$ can be determined
finally. In Fig.~\ref{tune-p3}, we just tune $p_3$ with  $f_{41}$
fixed to try to find the exact $z_h$ numerically. Recall that $z_h$
satisfies $f_b(z_h)=0$ and $\text{Test}(z_h)=0$. That is to say
$p_3$ and $f_{41}$ are not independent and there are relations among
them and $z_h$. This is very important in studying entanglement
temperature. Fig.~\ref{tune-p3} shows that how to find $z_h$ and
$p_3$ with
 $f_{41}$ fixed. In Fig.~\ref{tune-p3}(a), one can vary $p_3$ with
 $f_{41}$ fixed and find that the blue solid line and the blue dashed
line cross the same point in $x$ axis which means horizon has been
found. In Fig.~\ref{tune-p3}(b), we just show $\text{Test}(z)$
explicitly to confirm that there is one $z_h$ such that
$\text{Test}(z_h)=f_{b1}(z_h)=0$ for each $p_3$.

To closed this section, we should summarize the algorithm. Firstly,
one should figure out the series expansion of unknown function
$\phi_{b1}(z), A_{Et1}(z), f_{b1}(z)$ with asymptotical AdS boundary
condition. Here asymptotical AdS boundary condition can fix 2
integral constants. Secondly, one can find three integral constants
determined by IR boundary conditions and $\text{Test}(z)$. Finally,
finding the horizon position and one of integral constants with
shooting method will fix the final two integral constants. After
these three steps, all integral constants can be fixed numerically
and one can put them into the Einstein equations to produce
numerical black hole solution in this system.

\begin{figure}[h]
\begin{center}
\epsfxsize=6.5 cm \epsfysize=6.5 cm \epsfbox{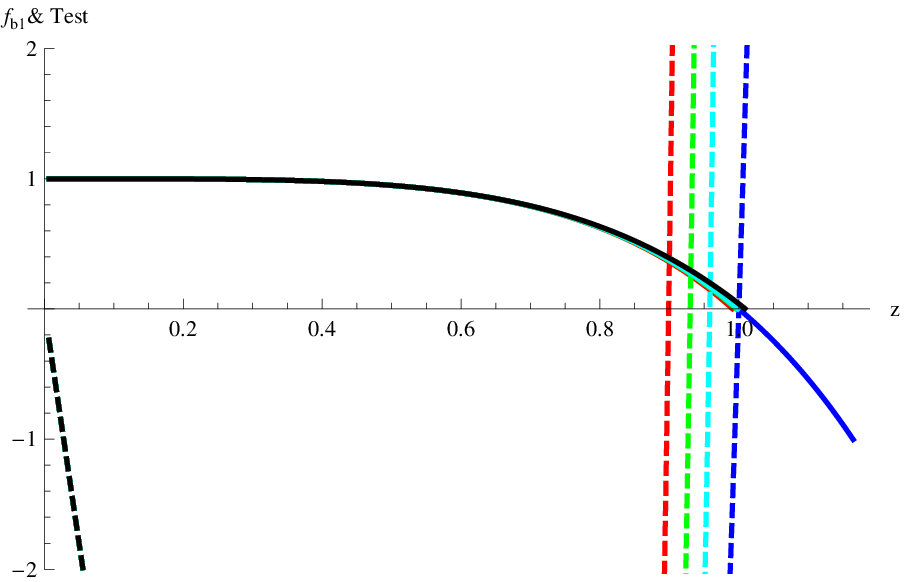}
\hspace*{0.1cm} \epsfxsize=6.5 cm \epsfysize=6.5
cm\epsfbox{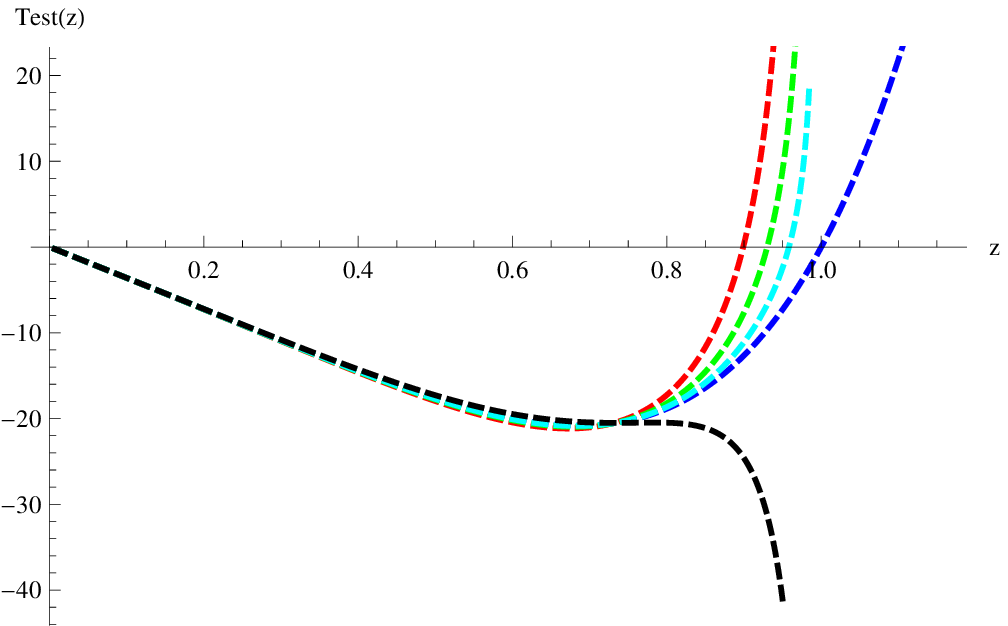} \vskip -0.05cm \hskip 0.15 cm
\textbf{( a ) } \hskip 6.5 cm \textbf{( b )} \\
\end{center}
\caption{The behavior of $f_{b1}(z)$ and $\text{Test}(z)$ when $p_3$
varies and $p_1=1.5,f_{41}=0.75$. The solid lines and dashed lines
stands for $f_{b1}(z)$ and $\text{Test}(z)$ respectively. And the
Red, Green, Cyan, Blue, Black line stand for
$p_3=-0.1,-0.15,-0.175,-0.18697,-0.3$ respectively.} \label{tune-p3}
\end{figure}

\subsection{The second numerical black hole solution }
Following the arithmetic given in subsection A.1, we would like to
find the second numerical black hole with potential
$V_{Et2}=-\frac{27}{4 L^2}-\frac{21 \cosh \left(\frac{2 \sqrt{2}
\phi}{3}\right)}{4 L^2}$. Here one should note that the powers order
of $z$ is not integer anymore, since they are constrained by
Einstein equation. Here we do not repeat numerical analysis
procedure as in the previous section. \footnote{If reader would like
to repeat the above analysis mentioned in (A.1), you can only
replace all the subscript index $b1$ of these functions with $b2$.}
We just choose various parameters to obtain the corresponding black
hole solution in the same way. Here we follow the same steps to find
the exact $z_h$ shown in Fig.~\ref{tune-p3-4}. In
Fig.~\ref{tune-p3-4}, one can vary $p_{7\over 2}$ with fixing
$f_{42}$ and find that the blue solid line and the blue dashed line
cross the same point in $x$ axis which means $z_h$ has been found.
Here the blue solid line and the blue dashed line stands for
$\text{Test}(z)$ and $f_{b2}(z)$ respectively. Where
$\text{Test}(z)$ defined by (\ref{test}) with replacing subscript
index $b1$ with $b2$. In Fig.~\ref{tune-p3-4}(b), we just show
$\text{Test}(z)$ explicitly to confirm that there is one $z_h$ such
that $\text{Test}(z_h)=f_{b2}(z_h)=0$ for each $p_{7\over 2}$. Once
$z_h$ is fixed with given $p_{1\over 2}$, the black hole solution
can be obtained numerically.
\begin{figure}[h]
\begin{center}
\epsfxsize=6.5 cm \epsfysize=6.5 cm \epsfbox{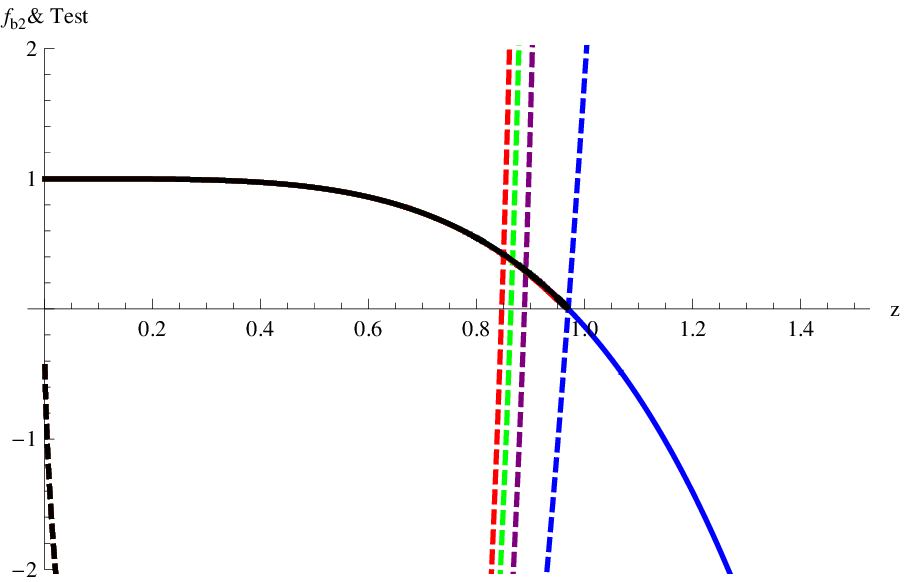}
\hspace*{0.1cm} \epsfxsize=6.5 cm \epsfysize=6.5
cm\epsfbox{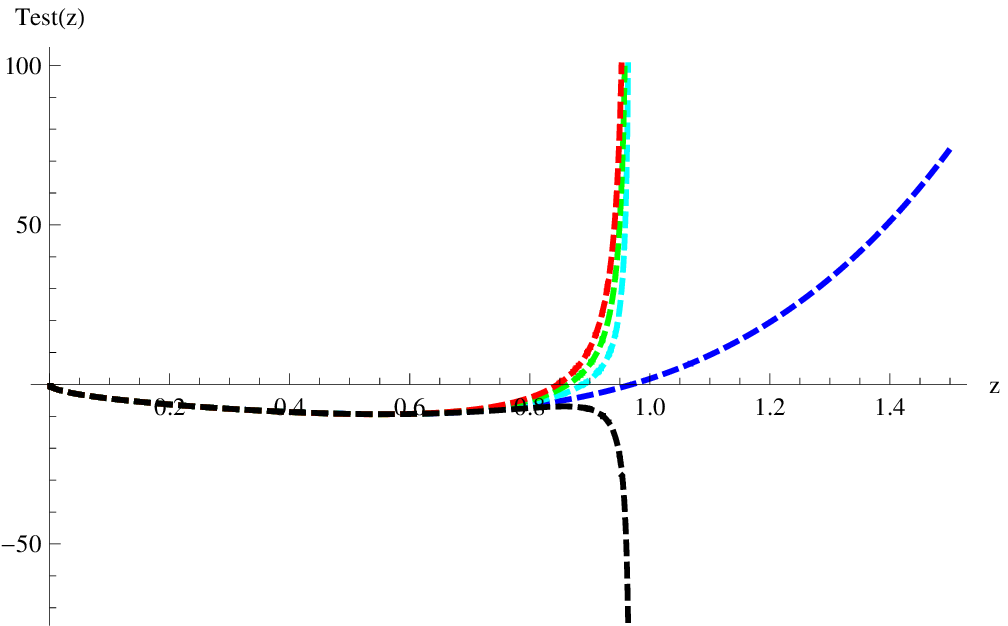} \vskip -0.05cm \hskip 0.15 cm
\textbf{( a ) } \hskip 6.5 cm \textbf{( b )} \\
\end{center}
\caption{The behavior of $f_{b2}(z)$ and $\text{Test}(z)$ when
$p_{7\over 2}$ varies and $p_{1\over 2}=1,f_{42}=1$. The solid lines
and dashed lines stands for $f_{b2}(z)$ and $\text{Test}(z)$
respectively. The Red, Green, Cyan, Blue, Black line stand for
$p_{7\over 2}=-0.01,-0.03,-0.05,-0.0706084,-0.09$ respectively.}
\label{tune-p3-4}
\end{figure}

\section{Other Analytic Solutions}

In this subsection, we would like to list other analytic solutions
generated by potential reconstruction approach in this paper. At
this stage, we have not studied related properties of these
solutions. It is interesting to study these solutions with
asymptotical AdS boundary condition from holographical point of view
in the future. Here we just only list other 4 solutions of ED system
(\ref{SGD}) with following ansatz \begin{equation}
\label{Ametric-Einsteinframe} ds_E^2 =\frac{{L^2}
e^{2A_{en}}}{z^2}\left(-f_{n}(z)dt^2
+\frac{dz^2}{f_{n}(z)}+dx^{i}dx^{i}\right).
\end{equation} Where $n$ denotes different solutions.

The 3rd solution is
\begin{eqnarray}
A_{e3}(z)&&=\log(1-\mu_3 z)\\
\phi_{3}(z)&&=-3\sqrt{2}\arctan(\sqrt{\mu_3 z})\\
f_{3}(z)&&=1+f_{43}\left(-\frac{3}{2}+(1-\mu_3 z)+\frac{3}{1-\mu_3 z}-\frac{1}{2(1-\mu_3 z)}+3\log(1-\mu_3 z)\right)\\
V_{E3}(\phi_3)&&=-\frac{1}{32 L^2}3 \cosh ^4(\frac{\phi_3 }{3 \sqrt{2}}) \\
&&\big( f_{43}\cosh \left(\sqrt{2} \phi_3 \right)+168 f_{43} \log \left(\text{sech}^2\left(\frac{\phi_3 }{3 \sqrt{2}}\right)\right)\\
&&+\cosh \left(\frac{2 \sqrt{2} \phi_3 }{3}\right) \left(24 f_{43} \log \left(\text{sech}^2\left(\frac{\phi_3}{3 \sqrt{2}}\right)\right)+42 f_{43}+8\right)\\
&&+\cosh \left(\frac{\sqrt{2} \phi_3 }{3}\right) \left(192 f_{43}
\log \left(\text{sech}^2\left(\frac{\phi_3 }{3
\sqrt{2}}\right)\right)+15 f_{43}+64\right)\nonumber\\&-&58
f_{43}+56\big)
\end{eqnarray}
Where $\mu_3, f_{43}$ are integral constants and $L$ stands for
asymptotical AdS radius.

The 4th solution can be expressed
\begin{eqnarray}
A_{e4}(z)&=&-\log(1+\frac{\mu_4^2z^2}{3(1+\mu_4 z)})\\
\phi_{4}(z)&=&\frac{3\sqrt{2}}{2}\log(1+\mu_4 z)\\
f_{4}(z)&=&1-f_{44}\frac{14-81(1+\mu_4 z)^2+84(1+\mu_4 z)^3-21(1+\mu_4 z)^6+4(1+\mu_4 z)^9}{756(1+\mu_4 z)^2}\\
V_{E4}(\phi_4)&=&-\frac{14 e^{\frac{2 \sqrt{2} \phi_4 }{3}}}{3
L^2}-\frac{20 e^{-\frac{1}{3} \left(\sqrt{2} \phi_4 \right)}}{3 L^2}
-\frac{2 e^{-\frac{1}{3} \left(4 \sqrt{2} \phi_4 \right)}}{3 L^2}+\\
&&\frac{f_{44}}{L^2} \Big(-\frac{5 e^{-\frac{1}{3} \left(\sqrt{2}
\phi_4 \right)}}{7}+\frac{2 e^{-\sqrt{2} \phi_4
}}{9}-\frac{e^{-\frac{1}{3} \left(4 \sqrt{2} \phi_4
\right)}}{14}+\frac{5 e^{\sqrt{2} \phi_4 }}{21}
\nonumber\\&-&\frac{e^{2 \sqrt{2} \phi_4 }}{126}-\frac{e^{\frac{2
\sqrt{2} \phi_4 }{3}}}{2}+\frac{5}{6}\Big),
\end{eqnarray}
where $\mu_4, f_{44}$ are integral constants and $L$ stands for
asymptotical AdS radius.

The 5th solution is
\begin{eqnarray}
A_{e5}(z)&=&-\log(\frac{3(-1+(1+\mu_5 z)^{5/3})}{5\mu_5 z(1+\mu_5 z)^{1/3}})\\
\phi_{5}(z)&=&\log(1+\mu_5 z)\\
f_{5}(z)&=&1-f_{45}\Big( \frac{9}{5}(1+\mu_5 z)^{5/3}-\frac{9}{10}(1+\mu_5 z)^{10/3}\nonumber\\&&+\frac{1}{5}(1+\mu_5 z)^5-\log(1+\mu_5 z)-\frac{11}{10}\Big)\\
V_{E5}(\phi_{5})&=&-\frac{3 e^{-\frac{8 \phi_{5} }{3}} \left(40
e^{\frac{5 \phi_{5} }{3}}+60 e^{\frac{10 \phi_{5} }{3}}\right)}{25
L^2}\nonumber\\&-&\frac{3 f_{45} e^{-\frac{8 \phi_{5} }{3}} \left(12
e^{\frac{10 \phi_{5} }{3}} (5 \phi_{5} -3)+e^{\frac{20 \phi_{5}
}{3}}-12 e^{5 \phi_{5} }+e^{\frac{5 \phi_{5} }{3}} (40 \phi_{5}
+44)+3\right)}{25 L^2},
\end{eqnarray}
Where $\mu_5, f_{45}$ are integral constants and $L$ stands for
asymptotical AdS radius.

The 6th solution is
\begin{eqnarray}
A_{e6}(z)&=&-\log(1+\mu_6^\alpha z^\alpha)\\
\phi_{6}(z)&=&3\sqrt{\frac{1+\alpha}{\alpha}}\arcsin(\sqrt{\mu_6^\alpha z^\alpha})\\
f_{6}(z)&=&1-f_{46} \mu_6^4
z^4\left(1+\frac{12\mu_6^{\alpha}z^{\alpha}}{4+\alpha}
+\frac{6\mu_6^{2\alpha}z^{2\alpha}}{2+\alpha}+\frac{4\mu_6^{3\alpha}z^{3\alpha}}{4+3\alpha}\right)\\
V_{E6}(\phi_{6})&=&\frac{3 \left(-256 \left(3 \alpha ^3+22 \alpha
^2+48 \alpha +32\right)\right)}{64 (\alpha +2) (\alpha +4) (3 \alpha
+4) L^2}\nonumber\\&+&\frac{3 \left(-32 \left(3 \alpha ^4+25 \alpha
^3+70 \alpha ^2+80 \alpha +32\right)\right)}{64 (\alpha +2) (\alpha
+4) (3 \alpha +4) L^2}\nonumber\\&&\sinh ^2\left(\frac{\alpha
\phi_{6} }{3 \sqrt{\alpha  (\alpha +1)}}\right) \left((3 \alpha +4)
\cosh \left(\frac{2 \alpha  \phi_{6} }{3 \sqrt{\alpha  (\alpha
+1)}}\right)-5 \alpha +12\right)\nonumber\\&+&\frac{3 \alpha ^2
\left(-120 \alpha ^3-134 \alpha ^2-114 \alpha -40\right) f_{46}
\sinh ^{\frac{8}{\alpha }+2}\left(\frac{\alpha  \phi_{6} }{3
\sqrt{\alpha (\alpha +1)}}\right)}{64 (\alpha +2) (\alpha +4) (3
\alpha +4) L^2}\nonumber\\&+&3 \alpha ^2 f_{46} \sinh
^{\frac{8}{\alpha }+2}\left(\frac{\alpha  \phi_{6} }{3 \sqrt{\alpha
(\alpha +1)}}\right)\frac{ \left(3 \left(24 \alpha ^3-35 \alpha
^2-49 \alpha -20\right) \cosh \left(\frac{2 \alpha  \phi_{6} }{3
\sqrt{\alpha (\alpha +1)}}\right)\right)}{64 (\alpha +2) (\alpha +4)
(3 \alpha +4) L^2}\nonumber\\&+&\frac{3 \alpha ^2 f_{46} \sinh
^{\frac{8}{\alpha }+2}\left(\frac{\alpha  \phi_{6} }{3 \sqrt{\alpha
(\alpha +1)}}\right) \left(6 \left(5 \alpha ^2-5 \alpha -4\right)
\cosh \left(\frac{4 \alpha  \phi_{6} }{3 \sqrt{\alpha  (\alpha
+1)}}\right)-4 \cosh \left(\frac{2 \alpha  \phi_{6} }{\sqrt{\alpha
(\alpha +1)}}\right)\right)}{64 (\alpha +2) (\alpha +4) (3 \alpha
+4) L^2}\nonumber\\&+&\frac{3 \alpha ^2 f_{46} \sinh
^{\frac{8}{\alpha }+2}\left(\frac{\alpha  \phi_{6} }{3 \sqrt{\alpha
(\alpha +1)}}\right) \left(\alpha ^2 \cosh \left(\frac{2 \alpha
\phi_{6} }{\sqrt{\alpha (\alpha +1)}}\right)+3 \alpha  \cosh
\left(\frac{2 \alpha  \phi_{6} }{\sqrt{\alpha  (\alpha
+1)}}\right)\right)}{64 (\alpha +2) (\alpha +4) (3 \alpha +4) L^2},
\end{eqnarray}
Where $\mu_6, f_{46}$ are integral constants and $L$ stands for
asymptotical AdS radius. In this case, one can deform this solution
by turning  the parameter $\alpha$ which is useful to build up
some asymptotical AdS background.

To close this subsection, we would like to add the following
comments. One can use potential reconstruction approach to generate
various gravity background in conformal ansatz and domain wall
ansatz.  One should note from the solutions given
 above that the geometric parameters contribute
to dilaton potential $V_{E1,E2,E3,E4,E5,E6}$. Changing these
parameters in the potential $V_{En}$ with $n=1,...,6$ means that the
theory is changed. In other words, different values of the
parameters $f_{4n},\mu_{n}$ in $V_{En}$ correspond to different
gravity theories. In some gravity solutions
\cite{Li:2011hp}\cite{He:2010ye} reconstructed by this method, it
seems inevitable that these different theories can be connected by
the same form of action with different values of parameters in
$V_{En}$. The different values of parameters corresponds to
different configuration of bulk field and different potential,
therefore, these theory are not equivalent to each other any more.
Once one constructs gravity background, for the stability of the
system, one should confirm the potential
$V_{E1},V_{E2},V_{E3},V_{E4},V_{E5},V_{E6}$ and
$A_{e1,e2,e3,e4,e5,e6}(z), f_{1,2,3,4,5,6}(z),
\phi_{1,2,3,4,5,6}(z)$ should satisfy the constrains from other
perspectives, e.g., Breitenlohner-Freedman bound of scalar field
near AdS boundary
\cite{Breitenlohner:1982bm}\cite{Breitenlohner:1982jf}, that the
total action is finite, well-defined boundary conditions of the
system and so on. Generally speaking, the method is efficient and
effective and using the approach needs ones to do something more to
make the solution self-consistently. In order to avoid arbitrary
dilaton potential generated by potential reconstruction, we arrange
a systematic method to obtain zero temperature solution and the
corresponding numerical black hole solution with the same dilation
potential in ED system. By following the logic present in this
paper, one can produce nontrivial thermal gas solutions and obtain
the black hole solutions numerically. This approach is not the one
from first principle (i. e. top-down) but an effective and efficient
way which shed light on study of gauge/gravity duality.


\begin{thebibliography}{99}
\bibitem{Maldacena:1997re}
  J.~M.~Maldacena,
  ``The large N limit of superconformal field theories and supergravity,''
  Adv.\ Theor.\ Math.\ Phys.\  {\bf 2}, 231 (1998)
  [Int.\ J.\ Theor.\ Phys.\  {\bf 38}, 1113 (1999)]  [arXiv:hep-th/9711200].
\bibitem{Gubser:1998bc}
  S.~S.~Gubser, I.~R.~Klebanov and A.~M.~Polyakov,
  ``Gauge theory correlators from non-critical string theory,''
  Phys.\ Lett.\  B {\bf 428}, 105 (1998)
  [arXiv:hep-th/9802109].
\bibitem{Witten:1998qj}
  E.~Witten,
  ``Anti-de Sitter space and holography,''
  Adv.\ Theor.\ Math.\ Phys.\  {\bf 2}, 253 (1998)
  [arXiv:hep-th/9802150].
\bibitem{Aharony:1999ti}
  O.~Aharony, S.~S.~Gubser, J.~M.~Maldacena, H.~Ooguri and Y.~Oz,
  ``Large N field theories, string theory and gravity,''
  Phys.\ Rept.\  {\bf 323}, 183 (2000)
  [arXiv:hep-th/9905111].

\bibitem{Ryu:2006bv}
  S.~Ryu and T.~Takayanagi,
  ``Holographic derivation of entanglement entropy from AdS/CFT,''
  Phys.\ Rev.\ Lett.\  {\bf 96}, 181602 (2006)
  [hep-th/0603001].


  \bibitem{Lewkowycz:2013nqa}
  A.~Lewkowycz and J.~Maldacena,
  ``Generalized gravitational entropy,''
  arXiv:1304.4926 [hep-th].

\bibitem{Casini:2011kv}
  H.~Casini, M.~Huerta and R.~C.~Myers,
  ``Towards a derivation of holographic entanglement entropy,''
  JHEP {\bf 1105}, 036 (2011)
  [arXiv:1102.0440 [hep-th]].


\bibitem{Headrick:2010zt}
  M.~Headrick,
  ``Entanglement Renyi entropies in holographic theories,''
  Phys.\ Rev.\ D {\bf 82}, 126010 (2010)
  [arXiv:1006.0047 [hep-th]].

\bibitem{Hartman:2013mia}
  T.~Hartman,
  ``Entanglement Entropy at Large Central Charge,''
  arXiv:1303.6955 [hep-th].

\bibitem{Faulkner:2013yia}
  T.~Faulkner,
  ``The Entanglement Renyi Entropies of Disjoint Intervals in AdS/CFT,''
  arXiv:1303.7221 [hep-th].



\bibitem{Nishioka:2009un}
  T.~Nishioka, S.~Ryu and T.~Takayanagi,
  `Holographic Entanglement Entropy: An Overview,''  J.\ Phys.\ A A {\bf 42}, 504008 (2009)  [arXiv:0905.0932 [hep-th]].

\bibitem{Takayanagi:2012kg}
  T.~Takayanagi,
  ``Entanglement Entropy from a Holographic Viewpoint,''
  Class.\ Quant.\ Grav.\  {\bf 29}, 153001 (2012)
  [arXiv:1204.2450 [gr-qc]].

\bibitem{Albash:2011nq}
  T.~Albash and C.~V.~Johnson,
  ``Holographic Entanglement Entropy and Renormalization Group Flow,''  JHEP {\bf 1202}, 095 (2012)  [arXiv:1110.1074 [hep-th]].

\bibitem{Myers:2012ed}
  R.~C.~Myers and A.~Singh,
  ``Comments on Holographic Entanglement Entropy and RG Flows,''  arXiv:1202.2068 [hep-th].

\bibitem{deBoer:2011wk}
  J.~de Boer, M.~Kulaxizi and A.~Parnachev,
  ``Holographic Entanglement Entropy in Lovelock Gravities,''  JHEP {\bf 1107}, 109 (2011)  [arXiv:1101.5781 [hep-th]].

\bibitem{Hung:2011xb}
  L.~-Y.~Hung, R.~C.~Myers and M.~Smolkin,
  ``On Holographic Entanglement Entropy and Higher Curvature Gravity,''  JHEP {\bf 1104}, 025 (2011)  [arXiv:1101.5813 [hep-th]].

\bibitem{Chen:2013qma}
  B.~Chen and J.~-j.~Zhang,
  ``Note on generalized gravitational entropy in Lovelock gravity,''
  arXiv:1305.6767 [hep-th].

\bibitem{Bhattacharyya:2013jma}
  A.~Bhattacharyya, A.~Kaviraj and A.~Sinha,
  ``Entanglement entropy in higher derivative holography,''
  arXiv:1305.6694 [hep-th].

\bibitem{Nishioka:2006gr}
  T.~Nishioka and T.~Takayanagi,
  ``AdS Bubbles, Entropy and Closed String Tachyons,''  JHEP {\bf 0701}, 090 (2007)  [hep-th/0611035].


\bibitem{Sun:2008uf}
  J.~-R.~Sun,
  ``Note on Chern-Simons Term Correction to Holographic Entanglement Entropy,''
  JHEP {\bf 0905}, 061 (2009)
  [arXiv:0810.0967 [hep-th]].


\bibitem{Klebanov:2007ws}
  I.~R.~Klebanov, D.~Kutasov and A.~Murugan,
  ``Entanglement as a probe of confinement,''  Nucl.\ Phys.\ B {\bf 796}, 274 (2008)  [arXiv:0709.2140 [hep-th]].

\bibitem{Pakman:2008ui}
  A.~Pakman and A.~Parnachev,
  ``Topological Entanglement Entropy and Holography,''  JHEP {\bf 0807}, 097 (2008)  [arXiv:0805.1891 [hep-th]].

\bibitem{Ogawa:2011fw}
  N.~Ogawa and T.~Takayanagi,
  ``Higher Derivative Corrections to Holographic Entanglement Entropy for AdS Solitons,''  JHEP {\bf 1110}, 147 (2011)  [arXiv:1107.4363 [hep-th]].  







\bibitem{Cai:2012sk}
  R.~-G.~Cai, S.~He, L.~Li and Y.~-L.~Zhang,
  ``Holographic Entanglement Entropy in Insulator/Superconductor Transition,''
  JHEP {\bf 1207}, 088 (2012)
  [arXiv:1203.6620 [hep-th]].

\bibitem{Cai:2012nm}
  R.~-G.~Cai, S.~He, L.~Li and Y.~-L.~Zhang,
  ``Holographic Entanglement Entropy on P-wave Superconductor Phase Transition,''
  JHEP {\bf 1207}, 027 (2012)
  [arXiv:1204.5962 [hep-th]].

\bibitem{Cai:2012es}
  R.~-G.~Cai, S.~He, L.~Li and L.~-F.~Li,
  ``Entanglement Entropy and Wilson Loop in St\'{u}ckelberg Holographic Insulator/Superconductor Model,''
  JHEP {\bf 1210}, 107 (2012)
  [arXiv:1209.1019 [hep-th]].

\bibitem{Nozaki:2013wia}
  M.~Nozaki, T.~Numasawa and T.~Takayanagi,
  ``Holographic Local Quenches and Entanglement Density,''
  arXiv:1302.5703 [hep-th].

\bibitem{Hartman:2013qma}
  T.~Hartman and J.~Maldacena,
  ``Time Evolution of Entanglement Entropy from Black Hole Interiors,''
  arXiv:1303.1080 [hep-th].

\bibitem{Nozaki:2013vta}
  M.~Nozaki, T.~Numasawa, A.~Prudenziati and T.~Takayanagi,
  ``Dynamics of Entanglement Entropy from Einstein Equation,''
  arXiv:1304.7100 [hep-th].

\bibitem{MBH}M.~B.~Hastings, ``An area law for
one-dimensional quantum systems'', J. Stat. Mech. (2007)
P08024[quant-ph/0705.2024]

\bibitem{Srednicki:1993im}
  M.~Srednicki,
  ``Entropy and area,''
  Phys.\ Rev.\ Lett.\  {\bf 71}, 666 (1993)
  [hep-th/9303048].

\bibitem{FMG}Francisco Castilho Alcaraz, Miguel Ibanez Berganza, German Sierra,
``Entanglement of low-energy excitations in Conformal Field
Theory,'' Phys.Rev.Lett.106:201601,2011[ arXiv:1101.2881[cond-mat]].

\bibitem{Masanes:2009tg}
  L.~Masanes,
  ``An Area law for the entropy of low-energy states,''
  Phys.\ Rev.\ A {\bf 80}, 052104 (2009)
  [arXiv:0907.4672 [quant-ph]].


\bibitem{Allahbakhshi:2013rda}
  D.~Allahbakhshi, M.~Alishahiha and A.~Naseh,
  ``Entanglement Thermodynamics,''
  arXiv:1305.2728 [hep-th].


\bibitem{Bhattacharya:2012mi}
  J.~Bhattacharya, M.~Nozaki, T.~Takayanagi and T.~Ugajin,
  ``Thermodynamical Property of Entanglement Entropy for Excited States,''
  Phys.\  Rev.\  Lett.\  110, {\bf 091602} (2013)
  [arXiv:1212.1164 [hep-th]].

\bibitem{Myers:2010xs}
  R.~C.~Myers and A.~Sinha,
  ``Seeing a c-theorem with holography,''
  Phys.\ Rev.\ D {\bf 82}, 046006 (2010)
  [arXiv:1006.1263 [hep-th]].


\bibitem{Guo:2013aca}
  W.~-Z.~Guo, S.~He and J.~Tao,
  ``Note on Entanglement Temperature for Low Thermal Excited States in Higher Derivative Gravity,''
  arXiv:1305.2682 [hep-th].




\bibitem{Fursaev:2006ih}
  D.~V.~Fursaev,
  ``Proof of the holographic formula for entanglement entropy,''
  JHEP {\bf 0609}, 018 (2006)
  [hep-th/0606184].


\bibitem{Barrella:2013wja}
  T.~Barrella, X.~Dong, S.~A.~Hartnoll and V.~L.~Martin,
  ``Holographic entanglement beyond classical gravity,''
  arXiv:1306.4682 [hep-th].

\bibitem{Faulkner:2013ana}
  T.~Faulkner, A.~Lewkowycz and J.~Maldacena,
  ``Quantum corrections to holographic entanglement entropy,''
  arXiv:1307.2892 [hep-th].


 \bibitem{Gubser-T}
S.~S.~Gubser and A.~Nellore,
  ``Mimicking the QCD equation of state with a dual black hole,''
  Phys.\ Rev.\  D {\bf 78}, 086007 (2008);
S.~S.~Gubser, A.~Nellore, S.~S.~Pufu and F.~D.~Rocha,
``Thermodynamics and bulk viscosity of approximate black hole duals
to finite temperature quantum chromodynamics,''
  Phys.\ Rev.\ Lett.\  {\bf 101}, 131601 (2008);
  S.~S.~Gubser, S.~S.~Pufu and F.~D.~Rocha,
  ``Bulk viscosity of strongly coupled plasmas with holographic duals,''
  JHEP {\bf 0808}, 085 (2008).

\bibitem{Gursoy-T}
U.~Gursoy, E.~Kiritsis, L.~Mazzanti and F.~Nitti, ``Deconfinement
and Gluon Plasma Dynamics in Improved Holographic QCD,'' Phys.\
Rev.\ Lett.\  {\bf 101}, 181601 (2008);
U.~Gursoy, E.~Kiritsis, G.~Michalogiorgakis and F.~Nitti, ``Thermal
Transport and Drag Force in Improved Holographic QCD,''
  JHEP {\bf 0912}, 056 (2009).



\bibitem{Farakos:2009fx}
  K.~Farakos, A.~P.~Kouretsis, P.~Pasipoularides,
  ``Anti de Sitter 5D black hole solutions with a self-interacting bulk scalar field: A Potential reconstruction approach,''
  Phys.\ Rev.\  {\bf D80}, 064020 (2009).
  [arXiv:0905.1345 [hep-th]].

\bibitem{Li:2011hp}
  D.~Li, S.~He, M.~Huang and Q.~S.~Yan,
  ``Thermodynamics of deformed AdS$_5$ model with a positive/negative quadratic
  correction in graviton-dilaton system,''
  JHEP {\bf 1109}, 041 (2011)
  [arXiv:1103.5389 [hep-th]].

\bibitem{Ohta:2009pe}
  N.~Ohta, T.~Torii,
  ``Black Holes in the Dilatonic Einstein-Gauss-Bonnet Theory in Various Dimensions IV: Topological Black Holes with and without Cosmological Term,''
  Prog.\ Theor.\ Phys.\  {\bf 122}, 1477-1500 (2009).
  [arXiv:0908.3918 [hep-th]].

\bibitem{Kolyvaris:2009pc}
  T.~Kolyvaris, G.~Koutsoumbas, E.~Papantonopoulos, G.~Siopsis,
  ``A New Class of Exact Hairy Black Hole Solutions,''
  Gen.\ Rel.\ Grav.\  {\bf 43}, 163-180 (2011).
  [arXiv:0911.1711 [hep-th]].


\bibitem{Cai:2012xh}
  R.~-G.~Cai, S.~He and D.~Li,
  ``A hQCD model and its phase diagram in Einstein-Maxwell-Dilaton system,''
  JHEP {\bf 1203}, 033 (2012)
  [arXiv:1201.0820 [hep-th]].


\bibitem{He:2011hw}
  S.~He, Y.~-P.~Hu and J.~-H.~Zhang,
  ``Hydrodynamics of a 5D Einstein-dilaton black hole solution and the corresponding BPS state,''
  JHEP {\bf 1112}, 078 (2011)
  [arXiv:1111.1374 [hep-th]].

\bibitem{He:2010ye}
  S.~He, M.~Huang, Q.~-S.~Yan,
  ``Logarithmic correction in the deformed $AdS_5$ model to produce the heavy quark potential and QCD beta function,''
  Phys.\ Rev.\  {\bf D83}, 045034 (2011).
  [arXiv:1004.1880 [hep-ph]].

\bibitem{Cai:2012eh}
  R.~-G.~Cai, S.~Chakrabortty, S.~He and L.~Li,
  ``Some aspects of QGP phase in a hQCD model,''
  JHEP {\bf 1302}, 068 (2013)
  [arXiv:1209.4512 [hep-th]].

\bibitem{KS}
  K.~Skenderis,
  ``Lecture notes on holographic renormalization,''
  Class.\ Quant.\ Grav.\  {\bf 19}, 5849 (2002)
  [hep-th/0209067].
\bibitem{SKS}
  S.~de Haro, S.~N.~Solodukhin and K.~Skenderis,
  ``Holographic reconstruction of space-time and renormalization in the AdS / CFT correspondence,''
  Commun.\ Math.\ Phys.\  {\bf 217}, 595 (2001)
  [hep-th/0002230].

\bibitem{Ryu:2006ef}
  S.~Ryu and T.~Takayanagi,
  ``Aspects of Holographic Entanglement Entropy,''
  JHEP {\bf 0608}, 045 (2006)
  [hep-th/0605073].



\bibitem{Narayan:2013qga}
  K.~Narayan,
  ``Non-conformal brane plane waves and entanglement entropy,''
  arXiv:1304.6697 [hep-th].




\bibitem{Gursoy:2008za}
  U.~Gursoy, E.~Kiritsis, L.~Mazzanti and F.~Nitti,
  ``Holography and Thermodynamics of 5D Dilaton-gravity,''
  JHEP {\bf 0905}, 033 (2009)
  [arXiv:0812.0792 [hep-th]].






\bibitem{Breitenlohner:1982bm}
  P.~Breitenlohner and D.~Z.~Freedman,
  ``Positive Energy in anti-De Sitter Backgrounds and Gauged Extended Supergravity,''
  Phys.\ Lett.\ B {\bf 115}, 197 (1982).

\bibitem{Breitenlohner:1982jf}
  P.~Breitenlohner and D.~Z.~Freedman,
  ``Stability in Gauged Extended Supergravity,''
  Annals Phys.\  {\bf 144}, 249 (1982).

\end{thebibliography}
\end{document}